%% file: usenixsecurity2026.tex
\documentclass[letterpaper,twocolumn,10pt]{article}
\usepackage{usenix}

\usepackage{cite}
\usepackage{amsmath,amssymb,amsfonts}

\usepackage{textcomp}
\usepackage{pifont}
\usepackage{makecell}
\usepackage{adjustbox}
\usepackage{algorithm}
\usepackage{algpseudocode}
\usepackage{acro}
\input{0-0-0-acronyms}

\usepackage{bm}
\usepackage{booktabs}
\usepackage{changepage}
\usepackage{colortbl}
\usepackage{comment}
\usepackage{dblfloatfix}
\usepackage{diagbox}
\usepackage{enumitem}
\usepackage{graphicx}
\usepackage{siunitx}
\usepackage[percent]{overpic}
\usepackage[dvipsnames]{xcolor}
\usepackage{xurl}
\usepackage{hyperref}
\hypersetup{
  colorlinks=true,
  linkcolor=blue,
  citecolor=red,
  urlcolor=blue
}

\usepackage{lipsum}
\usepackage{multirow}
\usepackage{pgfplots}
\pgfplotsset{compat=1.18}
\usepgfplotslibrary{polar}
\usepackage{pgfplotstable}
\usepackage{subcaption}
\usepackage{tabularx}
\usepackage{tikz}
\usetikzlibrary{shapes.geometric, arrows.meta, calc, positioning}
\usetikzlibrary{shapes, backgrounds, fit, shadows}
\usetikzlibrary{patterns}
\usepackage[most]{tcolorbox}
\usepackage{array}
\newcolumntype{L}[1]{<{\raggedright\arraybackslash}p{#1}}
\newcolumntype{Y}{<{\raggedright\arraybackslash}X}

\newcommand{\circlednum}[1]{
  \ifcase#1
  \or \ding{202}
  \or \ding{203}
  \or \ding{204}
  \or \ding{205}
  \or \ding{206}
  \or \ding{207}
  \or \ding{208}
  \or \ding{209}
  \or \ding{210}
  \or \ding{211}
  \else ?
  \fi
}

\def\BibTeX{{\rm B\kern-.05em{\sc i\kern-.025em b}\kern-.08em
T\kern-.1667em\lower.7ex\hbox{E}\kern-.125emX}}

\begin{document}


\title{\Large \bf Lights, Camera, Attack: Exploiting Temporal HDR Fusion with Pulsed Light}

\author{
{\rm Alkim Domeke}\\
Clemson University
\and
{\rm Michael Kühr}\\
Technical University of Munich
\and
{\rm Roman Gilliatt}\\
Clemson University
\and
{\rm Mohammad Hamad}\\
Technical University of Munich
\and
{\rm Sebastian Steinhorst}\\
Technical University of Munich
\and
{\rm Bing Li}\\
Clemson University
\and
{\rm Long Cheng}\\
Clemson University
\and
{\rm Mert D. Pesé}\\
Clemson University
}

\maketitle

\begingroup
\renewcommand{\thefootnote}{}
\footnotetext{DISTRIBUTION STATEMENT A. Approved for public release; distribution is unlimited. OPSEC11092}
\addtocounter{footnote}{-1}
\endgroup

\begin{abstract}
  \input{0-Abstract}
\end{abstract}

\section{Introduction}
\label{sec:1-Introduction}
\input{1-Introduction}

\section{Background}
\label{sec:4-Background}
\input{2-Background}

\section{Related Work}
\label{sec:2-Related_Work}
\input{2_3-Related_Work}

\section{Threat Model}
\label{sec:3-Threat_model}
\input{3-Threat_model}

\section{FLASH Attack Design}
\label{sec:4-Attack_Designs}

\input{4-Attack_Methodology}

\section{Experiments}
\label{sec:6-Experiments}
\input{6-Experiments}

\section{Proof-of-Concept Mitigation}
\label{sec:7-Defense}
\input{7-Defense}

\section{Discussion and Limitations}
\label{sec:8-Discussion_and_Limitations}
\input{8-Discussion_and_Limitations}

\section{Conclusion}
\label{sec:9-Conclusion}
\input{9-Conclusion}

\section*{Ethical Considerations}
\label{Ethical_Considerations}
\input{10-Ethical_appendix}

\section*{Open Science}

\label{Open_Science}
\input{10-Open_science_Appendix}

\section*{Acknowledgment}
This work was supported by Clemson University’s Virtual Prototyping of Autonomy Enabled Ground Systems (VIPR-GS), under Cooperative Agreement W56HZV-21-2-0001 with the US Army DEVCOM Ground Vehicle Systems Center (GVSC).

\bibliographystyle{plainurl}
\bibliography{references}

\appendix
\input{10-Appendix}

\end{document}

%% file: 0-0-0-acronyms.tex
\DeclareAcronym{flash}{
  short = \emph{FLASH} ,
  long  = Fusion-Level Attack by Saturating HDR
}
\DeclareAcronym{FLASH}{
  short = \emph{FLASH} ,
  long  = Fusion-Level Attack by Saturating HDR
}
\DeclareAcronym{AV}{
  short = AV ,
  long  = Autonomous Vehicle
}
\DeclareAcronym{AI}{
  short = AI ,
  long  = Artificial Intelligence
}
\DeclareAcronym{AVs}{
  short = AVs ,
  long  = Autonomous Vehicles
}
\DeclareAcronym{ISP}{
  short = ISP ,
  long  = Image Signal Processing
}
\DeclareAcronym{HDR}{
  short = HDR ,
  long  = High Dynamic Range
}
\DeclareAcronym{LDR}{
  short = LDR ,
  long  = Low Dynamic Range
}
\DeclareAcronym{ADAS}{
  short = ADAS ,
  long  = Advanced Driver-Assistance Systems
}
\DeclareAcronym{CCD}{
  short = CCD ,
  long  = Charge-Coupled Device
}
\DeclareAcronym{CMOS}{
  short = CMOS ,
  long  = Complementary Metal-Oxide-Semiconductor
}
\DeclareAcronym{AE}{
  short = AE ,
  long  = Auto-Exposure
}
\DeclareAcronym{AWB}{
  short = AWB ,
  long  = Automatic White Balance
}
\DeclareAcronym{FPM}{
  short = FPM ,
  long  = Flash Per Minute
}
\DeclareAcronym{fps}{
  short = fps ,
  long  = Frame Per Second
}
\DeclareAcronym{MUTCD}{
  short = MUTCD ,
  long  = Manual on Uniform Traffic Control Devices
}
\DeclareAcronym{ERS}{
    short = ERS, 
    long = Electronic Rolling Shutter
}
\DeclareAcronym{CAN}{
    short = CAN, 
    long = Controller Area Network
}
\DeclareAcronym{SIL}{
    short = SIL, 
    long = Software-in-the-Loop
}
\DeclareAcronym{IR}{
    short = IR, 
    long = Infrared
}
\DeclareAcronym{MAD}{
    short = MAD, 
    long = Mean Absolute Deviation
}

%% file: 0-Abstract.tex
Modern cameras widely use temporal High Dynamic Range (HDR) to improve visibility by capturing a sequence of exposures with different integration times and fusing them into a single image. This process implicitly assumes that scene illumination remains sufficiently stable during capture. We introduce FLASH (Fusion-Level Attack by Saturating HDR), an external pulsed-light attack that deliberately attacks this assumption by creating cross-exposure inconsistency before downstream perception. FLASH exploits an algorithmic assumption rather than relying on sensor damage or hardware failure, and requires neither physical camera access, access to raw exposure brackets, knowledge of the fusion algorithm, nor exact phase lock to the camera. Across eight physical camera platforms spanning embedded, surveillance, photography, smartphone, and automotive use cases, and matched optical controls, FLASH causes pipeline-dependent darkening, overexposure, and visibility loss. This includes extreme-darkening rates of 50.0\% on an iPhone 16 Pro and 33.7\% on a Wyze Battery Cam Pro. On the Wyze camera, FLASH triggers the system-level low-visibility response in 10/10 trials, compared with 0/10 continuous-light and randomized-frequency flashing controls. In a controlled stationary OpenPilot case study, 23.0\% of frames exhibit severe darkening in the traffic-cone target region, with target-background CNR decreasing by up to 90.8\%. Under FLASH, the OpenPilot interface also fails to display the system-level path state observed in the corresponding control trials. In a controlled night-only HDR reconstruction stress test, a proof-of-concept exposure-rejection defense reduces median output-brightness deviation by 79.16\%. These results show that temporal HDR fusion itself requires security-aware validation of exposure evidence.
\begin{figure}[h]
  \centering
  \includegraphics[width=0.45\textwidth]{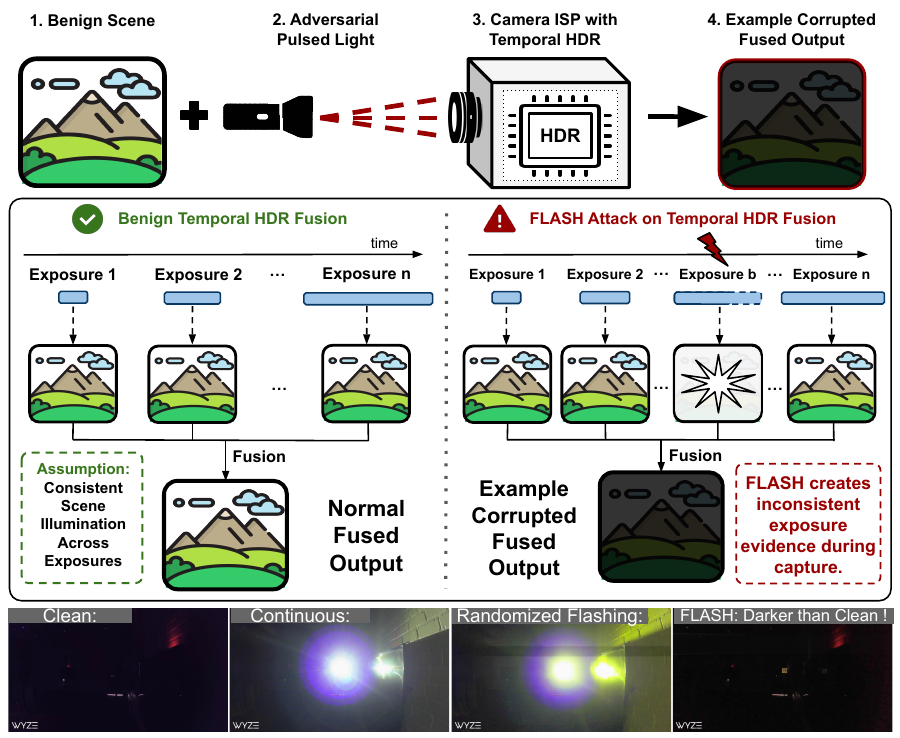}
  \caption{Overview of FLASH on temporal HDR fusion. Temporal HDR fuses exposures of different integration times under an assumption of stable scene illumination. FLASH creates cross-exposure inconsistency by perturbing one or more exposures, which can corrupt the fused output. The bottom row shows a Wyze Battery Cam Pro example, where continuous and randomized-frequency flashing brighten the scene while FLASH produces a darker output than the clean baseline.}
  \label{fig:teaser}
\end{figure}

%% file: 1-Introduction.tex
High Dynamic Range (HDR) imaging allows cameras to preserve visual detail across scenes containing both bright and dark regions. A common approach is temporal multi-exposure \ac{HDR}, in which the camera captures multiple observations of the same scene using different integration times and combines them into one output frame. Shorter exposures preserve information in bright regions, while longer exposures collect more light from darker regions; the camera's \ac{ISP} then aligns and fuses these observations into the final image~\cite{mertens2007exposure,debevec1997hdr}. Temporal multi-exposure fusion is identified in prior imaging literature as a widely applied \ac{HDR} technique~\cite{Wu:24} and is used in mobile computational photography~\cite{hasinoff2016burst}, commercial surveillance cameras~\cite{axis_wdr}, automotive imaging~\cite{takayanagi2022hdr}, and embedded vision systems~\cite{econ2021hdr,singh2023hdr}. Despite differences in implementation, these pipelines rely on a common assumption: the scene and its illumination remain sufficiently consistent while the sequence of exposures is captured~\cite{debevec1997hdr,hasinoff2016burst}.

This assumption creates a physical attack surface. If illumination changes rapidly during the exposure sequence, different exposures can record inconsistent versions of the same scene. Prior imaging work has shown that ordinary LED flicker can interfere with temporal \ac{HDR} capture~\cite{brading2016,sony_lfm,Deegan2018}. We show that this behavior can be induced deliberately. In particular, an external light source can inject different amounts of optical energy into different exposure intervals, causing the resulting exposure evidence to become mutually inconsistent. Rather than simply making the captured scene brighter, this inconsistency can interact with exposure fusion and subsequent \ac{ISP} processing to produce darkening, overexposure, contrast loss, or other corrupted output. The resulting failure therefore occurs during image formation, before any downstream perception model receives the frame.

We introduce Fusion-Level Attack by Saturating HDR (FLASH), a non-contact physical attack that deliberately creates this cross-exposure inconsistency using pulsed light. \ac{FLASH} specifically targets sequential temporal \ac{HDR} pipelines and does not target spatial or single-shot \ac{HDR} architectures that capture dynamic-range information simultaneously. The attacker requires neither physical access to the camera, access to raw exposure brackets, knowledge of the proprietary fusion algorithm, nor exact phase synchronization with the camera. Instead, periodic open-loop illumination creates unequal pulse overlap across exposure intervals. Longer integration windows generally provide greater opportunity for pulse overlap, but \ac{FLASH} does not require a particular number of exposures or a specific exposure index to be corrupted. This distinguishes \ac{FLASH} from continuous strong-light attacks that primarily disrupt acquisition through persistent sensor saturation~\cite{li2022adversarial,Petit2015}, as well as attacks that optimize perturbations against downstream perception models~\cite{goodfellow2015explaining,kurakin2017adversarial,eykholt2018robustphysicalworldattacksdeep}. \ac{FLASH} instead exploits an algorithmic assumption inside temporal \ac{HDR} fusion itself.

We evaluate \ac{FLASH} through analytical modeling, controlled simulation, matched optical controls, physical camera experiments, and a controlled stationary comma 3X/OpenPilot case study. Simulation characterizes how geometry, source strength, and ambient illumination affect the attack and shows that the effect is strongest under low ambient illumination, short distance, and near-axis aiming. Across physical embedded, surveillance, consumer, and smartphone camera pipelines, \ac{FLASH} produces device-dependent responses rather than a uniform brightness change. The strongest frame-level collapse occurs on the iPhone 16 Pro and Wyze Battery Cam Pro, with extreme-darkening rates of 50.0\% and 33.7\%, respectively, relative to continuous-light controls. On the Wyze camera, \ac{FLASH} also triggers the built-in low-visibility response in 10/10 trials, compared with 0/10 continuous-light and randomized-frequency flashing controls. In the controlled stationary comma 3X case study, 23.0\% of frames exhibit severe darkening in the traffic-cone target region, and target-background separability decreases by up to 90.8\%. Under \ac{FLASH}, the OpenPilot interface also displays neither the green planned path nor the red blocked-path state observed in the corresponding control conditions. We treat these as camera-stream degradation and displayed-interface observations and do not attribute them to a particular internal perception or planning component.

Because the vulnerability arises before exposure fusion, we also evaluate whether inconsistent exposure evidence can be rejected before reconstruction. In a controlled three-exposure night-only \ac{HDR} stress test, a proof-of-concept exposure-level rejection method detects the corrupted exposure in 756/756 cases with a 0.0\% clean false rejection rate and reduces median absolute luma deviation from 60.46\% to 12.60\%, a 79.16\% relative reduction. This defense evaluation is performed in a controllable reconstruction pipeline rather than inside proprietary commercial \ac{ISP} firmware.

This paper makes the following main contributions:
\begin{itemize}
    \item We identify temporal \ac{HDR} fusion as a physical attack surface and introduce \ac{FLASH}, which creates cross-exposure illumination inconsistency using externally generated pulsed light. The attack operates before downstream perception and does not require physical camera access, raw exposure access, knowledge of the fusion algorithm, or exact phase synchronization.

    \item We characterize \ac{FLASH} through analysis, simulation, matched optical controls, and physical evaluation across embedded, surveillance, consumer, smartphone, and automotive camera systems. The results show that different camera pipelines map cross-exposure inconsistency into different output failure modes, including frame-level collapse, brightening, exposure compensation, glare, and local visibility degradation. We further evaluate comma 3X camera-stream degradation and displayed OpenPilot path state in a controlled outdoor stationary setting.

    \item We develop and evaluate a proof-of-concept exposure-level rejection method that removes anomalous exposure evidence before fusion. In a controlled three-exposure night-only \ac{HDR} stress test, it rejects the corrupted exposure in 756/756 cases with a 0.0\% false rejection rate and reduces median absolute luma deviation by 79.16\%.
\end{itemize}

To facilitate reproducibility and support future research, we make our \ac{HDR}-Lab framework, defense method, and simulation files available as open-source code at \url{https://anonymous.4open.science/r/Lights-Camera-Attack-HDR-Manipulation-with-FLASH-Attacks-CB90/}.

%% file: 2-Background.tex
Modern camera systems use \acp{ISP} to convert raw sensor readings into images for downstream perception models. As reported by Kühr \textit{et al.}~\cite{kuehr2024isp}, \acp{ISP} usually include both hardware and software stages. Many of these pipelines are proprietary and not standardized, so they behave as black-box preprocessing before data reaches application-level algorithms.
\ac{HDR} and wide-dynamic-range processing are common components of modern camera pipelines, including automotive \ac{ISP} designs and image sensors with on-chip \ac{HDR} preprocessing~\cite{arm_mali_c71ae,omnivision_ox08b40}. These stages extend dynamic range in scenes with strong brightness variation, such as indoor-outdoor transitions, night photography, and backlit surveillance views. 
To achieve this, temporal \ac{HDR}-enabled cameras capture multiple exposures of the same scene, with different integration times, and combine them into one image. This principle follows early multi-exposure dynamic-range reconstruction work, where differently exposed pictures of the same scene are combined to recover a higher-dynamic-range representation~\cite{mann1995undigital,debevec1997hdr,mertens2007exposure}. In such pipelines, shorter exposures preserve highlight details, while longer exposures preserve shadow details.
However, temporal \ac{HDR} fusion assumes consistent illumination across bracketed exposures~\cite{debevec1997hdr,hasinoff2016burst}. In practice, flashing and other rapid lighting changes violate this condition and can degrade fusion quality because multi-exposure pipelines are designed for stable illumination~\cite{brading2016,sony_lfm,willassen20151280,Deegan2018}. Abrupt pulsed light can cause one or more exposures, particularly those with longer integration times, to receive substantially more injected light than others, creating cross-bracket inconsistency and potentially producing darkened fusion output.
\ac{HDR} controls vary across devices. Consumer cameras may not expose \ac{HDR} as a user setting, surveillance cameras often place it in administrative interfaces, and automotive systems may integrate it into sensor or \ac{ISP} configurations managed by the platform~\cite{apple_hdr_settings,samsung_camera_assistant,sony_auto_hdr,axis_wdr_settings,ti_adas_hdr_reference,omnivision_automotive_hdr}. Operational users may therefore be unable to disable \ac{HDR} or disabling it can reduce visibility in low-light, backlit, and high-contrast scenes~\cite{hasinoff2016burst}.

%% file: 2_3-Related_Work.tex
\noindent \textbf{Optical sensor attacks.}
Prior physical optical attacks commonly interfere with image acquisition or downstream perception. Light-based attacks use lasers, strobes, or infrared illumination to saturate sensors, introduce artifacts, or mislead perception models~\cite{fu2022remote,duan2021adversarial,zhou2018invisible,wang2021invisible,bhupathiraju2024vulnerability}. Continuous strong-light attacks cause broad frame-level overexposure through persistent illumination~\cite{li2022adversarial,Petit2015}, while spatial optical attacks place or project structured perturbations into selected scene or image regions~\cite{ghostimage_2020man,duan2021adversarial}. These attacks primarily target sensor capture or downstream perception rather than deliberately creating inconsistent exposure evidence inside temporal \ac{HDR} fusion.

\noindent \textbf{Timing-based camera attacks.}
Other attacks exploit temporal properties of camera capture. Rolling-shutter attacks modulate illumination timing to introduce adversarial stripes, erase objects, or spoof traffic-light signals by exploiting the sequential readout of image rows~\cite{sayles2021invisible,chen2021illumination,huang2022lights,kohler2022signal,yan2022rolling}. Scene-wide flashing attacks similarly vary illumination over time, but generally target downstream video-recognition models rather than the internal relationship among \ac{HDR} exposure brackets~\cite{Pony_2021_CVPR}. In contrast, \ac{FLASH} uses temporal illumination specifically to create unequal optical contamination across the exposure observations used for temporal \ac{HDR} fusion. It therefore targets exposure-fusion consistency rather than rolling-shutter readout or model-specific temporal sensitivity.

\begin{table*}[t]
    \centering
    \caption{Comparison of FLASH with prior physical optical attacks.}
    \vspace{-0.8em}
    \label{tab:related_work_comparison}
    \small
    \setlength{\tabcolsep}{3.5pt}
    \renewcommand{\arraystretch}{1.12}

    \begin{tabularx}{\textwidth}{@{}
        >{\raggedright\arraybackslash}X
        @{}
        >{\centering\arraybackslash}p{1.8cm}
        >{\centering\arraybackslash}p{2.3cm}
        >{\centering\arraybackslash}p{1.8cm}
        >{\centering\arraybackslash}p{1.8cm}
        >{\centering\arraybackslash}p{1.8cm}
        >{\centering\arraybackslash}p{1.8cm}}
        \toprule
        \makecell[l]{\textbf{Attack Class}} &
        \makecell[c]{\textbf{External}\\\textbf{Only}} &
        \makecell[c]{\textbf{No Camera-}\\\textbf{Specific Profiling}} &
        \makecell[c]{\textbf{Model}\\\textbf{Agnostic}} &
        \makecell[c]{\textbf{No Object}\\\textbf{Alignment}} &
        \makecell[c]{\textbf{No Exact}\\\textbf{Phase Lock}} &
        \makecell[c]{\textbf{Targets}\\\textbf{HDR Failure}} \\
        \midrule

        \rowcolor{gray!8}
        \makecell[l]{Camera blinding~\cite{Petit2015,fu2022remote}} &
        $\checkmark$ &
        $\checkmark$ &
        $\checkmark$ &
        $\checkmark$ &
        $\checkmark$ &
        $\times$ \\

        \makecell[l]{Infrared injection~\cite{wang2021invisible}} &
        $\checkmark$ &
        $\times$ &
        $\checkmark$ &
        \textit{$(\checkmark)$} &
        $\checkmark$ &
        $\times$ \\

        \rowcolor{gray!8}
        \makecell[l]{Scene-wide flashing~\cite{Pony_2021_CVPR}} &
        $\checkmark$ &
        $\checkmark$ &
        $\times$ &
        $\checkmark$ &
        $\checkmark$ &
        $\times$ \\

        \makecell[l]{Rolling-shutter attacks~\cite{sayles2021invisible,kohler2021they,yan2022rolling}} &
        $\checkmark$ &
        $\times$ &
        \textit{$(\checkmark)$} &
        \textit{$(\checkmark)$} &
        \textit{$(\checkmark)$} &
        $\times$ \\

        \rowcolor{gray!8}
        \makecell[l]{Spatial optical attacks~\cite{ghostimage_2020man,duan2021adversarial}} &
        $\checkmark$ &
        \textit{$(\checkmark)$} &
        $\times$ &
        $\times$ &
        $\checkmark$ &
        $\times$ \\

        \textit{FLASH} &
        $\checkmark$ &
        $\checkmark$ &
        $\checkmark$ &
        $\checkmark$ &
        $\textit{$(\checkmark)$}^{\dagger}$ &
        $\checkmark$ \\

        \bottomrule
    \end{tabularx}

    \vspace{0.2em}
    \begin{minipage}{0.99\textwidth}
        \footnotesize
        \textit{Terms.}
        A checkmark indicates that the property holds, a cross that it does not, and a parenthesized checkmark that it varies across the cited attacks. "External only" means no access to the camera, ISP, or host system.
        "No camera-specific profiling" means no target-specific camera characterization.
        "Model agnostic" means no downstream model access or optimization.
        "No object alignment" means no pattern placement on a scene object or image region.
        "No phase lock" means no exact synchronization with sensor readout or exposure timing.
        Line of sight and sufficient optical power remain necessary.

        $\dagger$ FLASH still requires a pulse cadence that overlaps vulnerable exposure brackets.
    \end{minipage}
\end{table*}

\noindent \textbf{ISP and HDR failures.}
Prior work has also studied attacks and non-adversarial failures within camera-processing pipelines. Pipeline attacks inject faults or manipulate stages such as sensor interfaces, \ac{ISP} parameters, and image-scaling operations~\cite{oyama2024adversarial,phan2021adversarial,li2022adversarial,li2024image,xiao2019seeing}. Separately, the imaging literature documents failures of temporal \ac{HDR} under changing illumination. LED flicker can cause different exposure intervals to capture different illumination states~\cite{brading2016,sony_lfm,Deegan2018}. When exposure timing does not align with an LED's illumination cycle, the source may be captured inconsistently across observations~\cite{sony_lfm,Deegan2018}. Longer integration times and multi-exposure capture have been used to reduce such flashing artifacts by increasing the likelihood that illumination is observed during capture~\cite{sony_lfm,Deegan2018}. Other known \ac{HDR} failures include ghosting caused by scene motion~\cite{xiao2022deep} and halos or posterization introduced during tone mapping~\cite{hore2014halo}. Emerging spatial \ac{HDR} architectures, including split-diode, sub-pixel, and multi-tap designs, reduce the temporal exposure gap by capturing dynamic-range information concurrently rather than through sequential exposure brackets~\cite{willassen20151280,SubPixel2018,MultiTap2025}; these architectures are therefore outside the scope of this work.

Prior work thus establishes that temporal illumination variation can degrade \ac{HDR} imaging, but does not formulate deliberate cross-exposure illumination inconsistency as a physical attack on temporal multi-exposure fusion. \ac{FLASH} targets this fusion assumption inside the camera \ac{ISP}, rather than relying on persistent sensor saturation, rolling-shutter readout, or optimization against a downstream perception model.

%% file: 3-Threat_model.tex
\noindent \textbf{System Model.}
We model the victim as a camera that forms temporal \ac{HDR} output by capturing \(n \geq 2\) exposures with different integration times and combining them through an \ac{ISP} fusion and tone-mapping pipeline~\cite{chaudhari2021merging}. Temporal \ac{HDR} assumes that scene illumination remains sufficiently consistent across these exposure observations, as detailed in Appendix~\ref{Appendix:math_analysis}. \ac{FLASH} violates this assumption by injecting pulsed light during capture so that one or more exposures receive unequal injected illumination. Longer exposures generally provide greater opportunity for pulse overlap, but the attack does not require a particular exposure index to be corrupted. The resulting cross-exposure inconsistency can produce darkened, overexposed, clipped, or otherwise distorted \ac{HDR} output.

\noindent \textbf{Attacker Goal.}
The attacker aims to corrupt the final camera output by inducing inconsistent exposure evidence during temporal \ac{HDR} capture. The goal is to produce abnormal brightness, contrast, or visibility before the frame reaches downstream human or machine perception. We treat \ac{FLASH} as a camera-output attack and do not assume control over, or knowledge of, any downstream perception model.

\noindent \textbf{Attacker Capabilities.}
The attacker controls only an external pulsed-light source and can approximately position and aim it toward the target camera. The attack is non-contact, open-loop, and black-box. The attacker does not require physical access to the camera or host system, access to RAW exposure buffers, exposure registers, \ac{ISP} firmware or settings, sensor trigger lines, internal telemetry, or the downstream perception model. Exact knowledge of the exposure schedule, fusion algorithm, or camera frame rate is not assumed, and exact phase synchronization with sensor capture is not required. Public or externally observable information, such as approximate camera placement and the use of temporal multi-exposure \ac{HDR}, may be used to select a feasible attack configuration.

\noindent \textbf{Operational Constraints.}
\ac{FLASH} requires line of sight between the optical source and the target camera and sufficient received illumination to produce unequal contamination across exposure observations. Attack feasibility therefore depends on source strength, camera-to-source distance, aiming angle, relative source height, ambient illumination, and temporal pulse overlap. These factors constrain the effective operating region of the attack and are evaluated experimentally in Section~\ref{sec:6-Experiments}. The attack does not require continuous illumination or permanent modification of the environment. Successful execution requires only that the pulse train overlap the camera's exposure sequence unevenly enough to create cross-exposure inconsistency; the specific timing relationship is described in Section~\ref{sec:4-Attack_Designs}.

\noindent \textbf{Evaluation Scope.}
We evaluate \ac{FLASH} at three levels. First, controlled simulation isolates how spatial, illumination, and temporal conditions affect \ac{HDR} output degradation. Second, component-level physical experiments characterize the response of black-box embedded, surveillance, consumer, and smartphone camera pipelines using matched optical controls. Third, a controlled outdoor stationary comma 3X/OpenPilot case study evaluates camera-stream degradation and displayed path states using a speed-limit sign, pedestrian, traffic cone, and general road scene. These are evaluation settings rather than distinct attack mechanisms; all use the same external pulsed-light mechanism to create cross-exposure inconsistency before downstream perception. We additionally use the Raspberry Pi Camera Module 3, AI Camera, and HQ Camera as controllable component-level platforms for characterization, parameter sweeps, and baseline comparisons.
\newline

%% file: 4-Attack_Methodology.tex
\begin{figure}[tbp]
  \centering
  \includegraphics[width=0.46\textwidth]{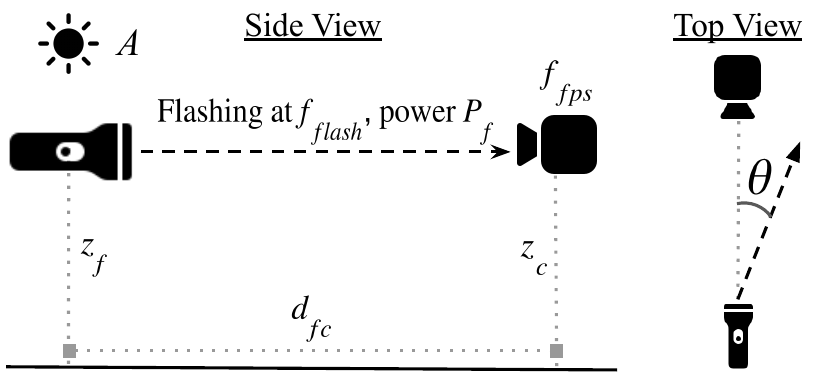}
  \caption{Physical setup and variables used in the FLASH evaluation. The side view shows the flashlight height $z_f$, camera height $z_c$, and horizontal camera-to-flashlight distance $d_{fc}$. The flashlight emits pulses with strength \(P_f\) and frequency \(f_{\mathrm{flash}}\), while the camera records at frame rate \(f_{\mathrm{fps}}\) under ambient-light condition \(A\). The top view defines the flashlight yaw angle $\theta$ between its optical axis and the line from the flashlight to the camera. At $\theta=0^\circ$, the flashlight points directly at the camera.}
  \label{fig:attack_parameters}
\end{figure}

\subsection{Overview}
\ac{FLASH} targets cameras that form each temporal \ac{HDR} output by fusing \(n \geq 2\) exposure observations captured with different integration times. Let $\mathcal{B}=\{1,\ldots,n\}$ denote the set of exposure indices, and let $T_b$ denote the integration time of the $b$th exposure, for $b\in\mathcal{B}$. The attacker directs periodic bright optical pulses toward the camera during capture. Because exposure windows differ in duration and temporal offset, the pulses can overlap different exposures by different amounts. Sufficiently unequal injected energy creates cross-exposure inconsistency that enters the proprietary fusion and tone-mapping pipeline and can produce darkening, overexposure, or visibility loss.

Longer exposure windows generally provide greater opportunity for pulse overlap, but \ac{FLASH} does not require a specific number of exposures or a particular exposure to be corrupted. The affected subset depends on the exposure schedule, pulse width, pulse frequency, and relative phase.
The attack proceeds by placing and aiming the source from a feasible line-of-sight location, selecting a source strength and pulse cadence, and inducing unequal contamination across exposure observations. \ac{FLASH} does not assume access to the exposure schedule, sensor readout, \ac{ISP}, or fusion algorithm.

\subsection{Attack Parameters}
\label{attack_parameters}
\noindent We distinguish attacker-selected settings from camera and environmental conditions. The attacker-selected configuration is $
\boldsymbol{a}
=
[z_f,\,d_{fc},\,\theta,\,P_f,\,f_{\mathrm{flash}}]^{\top},
$ while $\boldsymbol{s}
=
[z_c,\,A,\,f_{\mathrm{fps}}]^{\top}$ describes the operating condition. Camera height \(z_c\) is fixed by the platform, while \(A\) denotes the ambient-light condition and \(f_{\mathrm{fps}}\) the frame rate. We represent \(A\) as \(P_a^{\mathrm{sim}}\) in simulation and \(E_{v,a}^{\mathrm{phys}}\) in physical experiments.
For a given \(\boldsymbol{s}\), the attacker selects $\boldsymbol{a}\in\mathcal{C}(\boldsymbol{s})$, where \(\mathcal{C}(\boldsymbol{s})\) contains configurations satisfying source-range, line-of-sight, hardware, and safety constraints. Appendix~\ref{app:flash_constraints_interactions} formalizes these constraints. \ac{FLASH} does not solve a white-box optimization problem; we instead evaluate controlled sweeps over feasible configurations.

\noindent\textbf{Spatial and Optical Parameters.}
We use a right-handed coordinate system with a horizontal reference plane and vertical axis \(z\). In driving case studies, the reference plane is the road surface; in other settings, it is the local mounting plane. Figure~\ref{fig:attack_parameters} depicts this geometry.

The target camera is mounted at height \(z_c\). We define the effective vertical offset and source-to-camera distance as:
\[
z_f^{\mathrm{eff}}=z_f-z_c,
\qquad
r_{fc}=\sqrt{d_{fc}^{2}+\left(z_f-z_c\right)^{2}}.
\]

A first-order model of the flash irradiance received by the camera is:
\[
E_f(\boldsymbol{a},\boldsymbol{s})
\approx
\kappa
\frac{P_f g(\theta)}
     {d_{fc}^{2}+\left(z_f-z_c\right)^{2}},
\]
where \(g(\theta)\) represents the directional beam response and \(\kappa\) captures fixed source and optical factors. The actual camera response is measured empirically because the source optics, sensor response, and \ac{ISP} are treated as black boxes.

\noindent\textbf{Flashlight Height.}
Let \(z_f\in\mathbb{R}_{\ge 0}\) denote the flashlight height above the reference plane, with $z_f\in[z_{\min},z_{\max}]$.

\noindent\textbf{Camera-to-Flashlight Horizontal Distance.}
Let \(d_{fc}\in\mathbb{R}_{>0}\) denote the horizontal distance between the flashlight and camera, with $d_{fc}\in[d_{\min},d_{\max}]$.

\noindent\textbf{Flashlight Angle.}
Let \(\theta\in\mathbb{R}\) denote the yaw angle between the flashlight optical axis and the direction from the flashlight to the camera. Received irradiance is affected through \(g(\theta)\), with $\theta\in[\theta_{\min},\theta_{\max}]$.

\noindent\textbf{Ambient Illumination.}
We distinguish the ambient-light quantity used in simulation from that measured in physical experiments. In simulation, ambient lighting is controlled by area-light source power \(P_a^{\mathrm{sim}}\), while physical ambient illuminance is measured at the camera plane as \(E_{v,a}^{\mathrm{phys}}\) in lux. These quantities are not treated as directly equivalent.

\noindent\textbf{Flash Emission Strength.}
Let \(P_f\in\mathbb{R}_{\ge 0}\) denote the flashlight emission strength. In simulation, \(P_f\) denotes the nominal attack-source power in watts; in physical experiments, it denotes the corresponding source setting used by the hardware. It determines received flash irradiance together with distance, relative height, and aiming angle, with $P_f\in[P_{f,\min},P_{f,\max}]$.

\noindent\textbf{Temporal Interaction.}
\label{sec:temporal_parameters}
\begin{figure}[t]
    \centering
    \includegraphics[width=\linewidth]{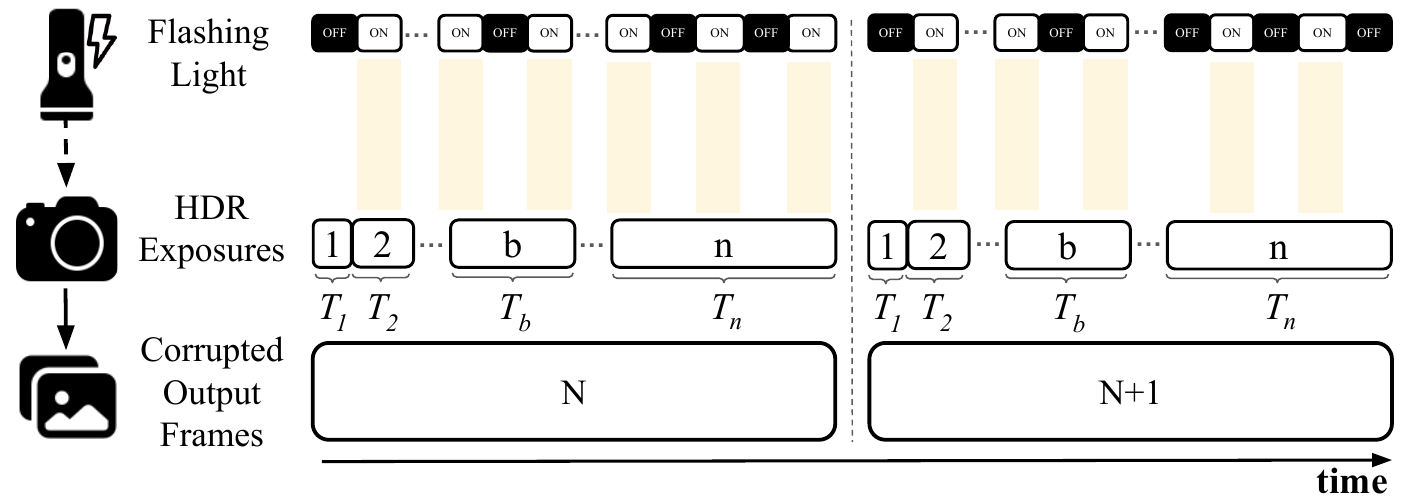}
    \caption{\textbf{Representative overview of \ac{FLASH} timing across consecutive HDR frames.} Each output frame \(N\) is formed by fusing \(n\) exposures with increasing integration times, \(T_1 < T_2 < \cdots < T_n\). Periodic light pulses overlap different subsets of these exposures, creating cross-exposure inconsistencies that corrupt the fused HDR output.}
    \label{fig:flash-timing}
\end{figure}
The temporal interaction depends on camera frame rate \(f_{\mathrm{fps}}\), flash frequency \(f_{\mathrm{flash}}\), pulse width \(\tau_p\), and relative phase \(\phi\). In our evaluation, \(f_{\mathrm{flash}}\) is attacker-selected, \(f_{\mathrm{fps}}\) is an operating condition, the pulse width is fixed by the source configuration, and no exact phase lock is used.
Temporal \ac{HDR} implementations acquire \(n \geq 2\) observations during distinct exposure windows. These may be sequential frames or staggered intervals within a frame cycle. For exposure \(b\in\mathcal{B}\), \(T_b>0\) denotes its integration time. The number, duration, and temporal offsets of the exposures depend on the camera implementation.

For staggered sensor \ac{HDR}, exposure intervals for a sensor row may be separated by readout offsets: Willassen \textit{et al.} report long and short exposures of \(43~\mathrm{ms}\) and \(0.043~\mathrm{ms}\)~\cite{willassen20151280}, while other commercial CMOS architectures report representative timings of \(T_S\approx0.8~\mathrm{ms}\), \(T_M\approx6.25~\mathrm{ms}\), and \(T_L\approx50.0~\mathrm{ms}\)~\cite{cvf_hdr_imaging}. These examples motivate modeling the sequence generically. For \(b,c\in\mathcal{B}\) with \(T_b>T_c\), the longer exposure provides greater opportunity for overlap under an unknown relative phase.

Let \(f_{\mathrm{fps}}\in\mathbb{R}_{>0}\) be the nominal camera frame rate, with output-frame start time:
\[
t_k=\frac{k}{f_{\mathrm{fps}}},
\qquad
f_{\mathrm{fps}}\in
[f_{\mathrm{fps},\min},f_{\mathrm{fps},\max}].
\]

Let \(f_{\mathrm{flash}}\in\mathbb{R}_{>0}\) denote pulse frequency, with
\[
T_{\mathrm{flash}}=\frac{1}{f_{\mathrm{flash}}},
\qquad
f_{\mathrm{flash}}\in
[f_{\mathrm{flash},\min},f_{\mathrm{flash},\max}].
\]

For exposure \(b\in\mathcal{B}\) associated with output frame \(k\), let $W_{k,b}=[t_k+\delta_b,\;t_k+\delta_b+T_b]$ denote its exposure interval, where \(\delta_b\) is its temporal offset. 

Let \(p(t;f_{\mathrm{flash}},\tau_p,\phi)\in[0,1]\) denote the normalized pulse train. The injected flash energy is proportional to
\[
Q_{k,b}
=
E_f(\boldsymbol{a},\boldsymbol{s})
\int_{W_{k,b}}
p(t;f_{\mathrm{flash}},\tau_p,\phi)\,dt.
\]

The spatial and optical settings determine \(E_f\), while frame rate, flash frequency, pulse width, relative phase, exposure duration, and exposure offset determine pulse overlap. \ac{FLASH} seeks configurations for which \(Q_{k,b}\neq Q_{k,c}\) for at least some \(b,c\in\mathcal{B}\), with sufficient difference to disturb temporal \ac{HDR} fusion.

Let \(X_{k,b}\) denote the clean measurement for exposure \(b\), and let \(\mathcal{H}\) denote the unknown temporal \ac{HDR} fusion and tone-mapping pipeline. The attacked and clean outputs are
\[
I_k^{\mathrm{FLASH}}
=
\mathcal{H}
\left(
\left\{
X_{k,b}+Q_{k,b}
\right\}_{b\in\mathcal{B}}
\right),
I_k^{\mathrm{clean}}
=
\mathcal{H}
\left(
\left\{
X_{k,b}
\right\}_{b\in\mathcal{B}}
\right).
\]

Because \(\mathcal{H}\), the exposure offsets, exact exposure durations, and number of exposure observations are generally unavailable, \ac{FLASH} does not compute an internal gradient or exact phase schedule. Section~\ref{sec:6-Experiments} instead measures the output effect through controlled parameter sweeps.

\noindent\textbf{Open-Loop Cadence Selection.}
\label{sec:fps_agnostic_timing}
For a stable nominal frame rate, selecting $\frac{f_{\mathrm{flash}}}{f_{\mathrm{fps}}}
\in\mathbb{Z}_{>0}$ causes the pulse pattern to repeat at the same nominal offset across output frames. This provides a repeatable cadence relationship but does not guarantee selective overlap with a particular exposure; actual overlap also depends on pulse width, relative phase, exposure durations, and offsets.

Our physical evaluation considers $F_{\mathrm{eval}}=\{24,30,60,120\}\ \mathrm{fps}$. Their least common multiple gives \(f_{\mathrm{flash}}=120~\mathrm{Hz}\), producing integer ratios of \(5\), \(4\), \(2\), and \(1\), respectively. The commercial flashlight controller reports cadence in \(\mathrm{RPM}\); under its cycles-per-minute convention, \(120~\mathrm{Hz}\) corresponds to \(7200~\mathrm{RPM}\).
For fractional or fluctuating rates such as \(29.97\) or \(59.94~\mathrm{fps}\), relative phase drifts over time. We therefore treat the common-multiple cadence as a nominal selection rather than a guarantee of exact synchronization. Its effectiveness is established empirically in the evaluation.

%% file: 6-Experiments.tex
We evaluate \ac{FLASH} in three stages. First, simulation characterizes the physical and environmental conditions under which \ac{FLASH} causes HDR output degradation. Second, physical camera experiments characterize how black-box camera pipelines respond relative to matched optical controls. Third, a controlled stationary comma 3X/OpenPilot case study evaluates camera-stream degradation and displayed path states.

\noindent\textbf{Evaluation Questions.}
\label{subsec:eval_questions_metrics}
In this section, we evaluate the following EQs:
\begin{itemize}[noitemsep, topsep=0pt]
    \item \textit{\textbf{EQ1}: Under what physical and environmental conditions does \ac{FLASH} cause camera-output degradation?}  We evaluate the effects of source distance, angle, height, strength, ambient-light condition, and capture frame rate through controlled simulation and physical parameter sweeps.
    \item \textit{\textbf{EQ2}: How do black-box physical camera pipelines respond to \ac{FLASH} relative to matched optical controls?} We compare \ac{FLASH} with clean, continuous-light, and randomized-frequency conditions across physical camera platforms and characterize the resulting response types using output-level luma and target-background contrast measurements.
    \item \textit{\textbf{EQ3}: In controlled stationary comma 3X/OpenPilot trials, how do the camera stream and displayed path state differ under \ac{FLASH}?} We separately evaluate offline camera-stream degradation and the path state displayed by the OpenPilot's own interface. We do not observe OpenPilot's internal perception or planning states.
\end{itemize}

\subsection{Evaluation Setup}
\label{sec:5-Evaluation_Setup}

\input{5-Evaluation_Setup}

\subsection{Simulation-Based Evaluation}
\label{subsec:simulation_eval}

The simulation evaluation addresses EQ1 by varying attack distance, horizontal angle, source height, optical power, and simulated ambient-light condition and measuring the resulting HDR output degradation after ISP processing. For each condition, we report a normalized output-degradation score relative to the matched no-attack baseline. Negative values indicate output darkening or compression, while positive values indicate brightening relative to the clean baseline. Scores at or below $-100\%$ indicate catastrophic \ac{HDR} failure, where the simulated \ac{ISP} output collapses to an extremely darkened frame. We interpret these cases as defined in Section~\ref{subsec:eval_questions_metrics}.

\begin{figure*}[t]
    \centering
    \includegraphics[width=\textwidth]{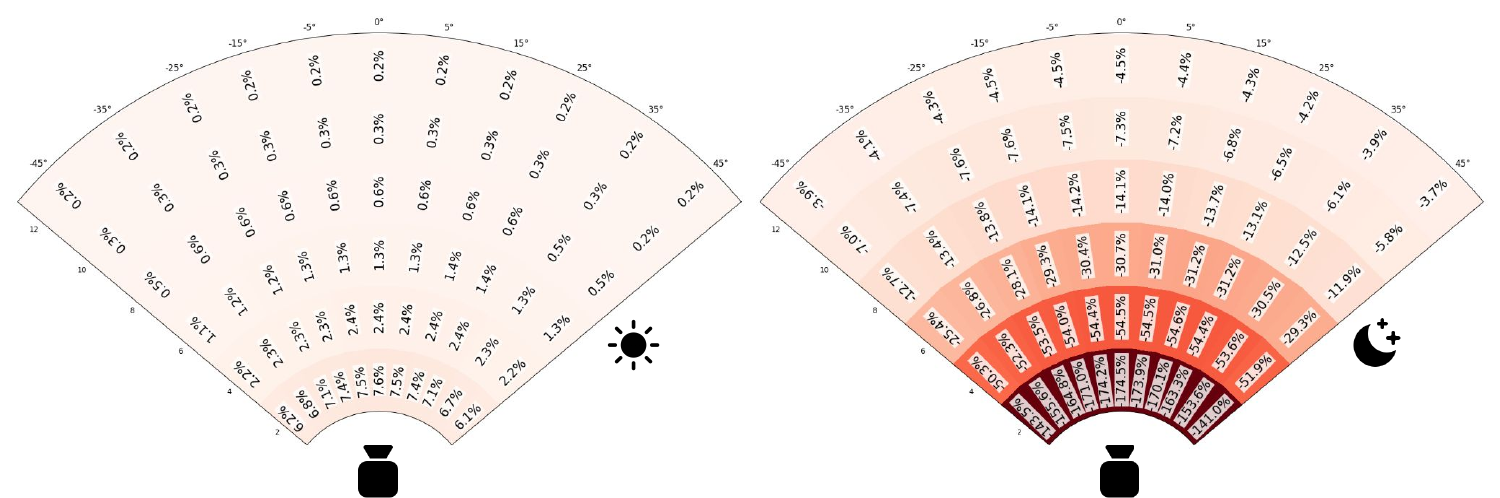}
    \caption{Simulation spatial sweep results for the flashlight attack. The source is fixed at \SI{12}{W} and \SI{1.5}{m}, while distance and horizontal aim angle are varied. The left plot shows the daytime baseline and the right plot shows the nighttime baseline.} 
    \label{fig:spatial_simulation_results}
\end{figure*}

\noindent \textbf{Spatial sweep.}
We first sweep distance ($d_{fc}$) and horizontal angle ($\theta$). The flashlight is fixed at \SI{12}{W} and \SI{1.5}{m}, while distance varies from \SI{2}{m} to \SI{20}{m} and aim angle varies from $-45^{\circ}$ to $45^{\circ}$. Figure~\ref{fig:spatial_simulation_results} shows that the strongest response occurs at short range and near $0^{\circ}$ aim. The effect drops sharply after approximately \SI{10}{m} and weakens as distance increases or the beam moves off axis. Nighttime results are substantially stronger, with extreme-darkening failures appearing only under low ambient illumination. Daytime results remain below about $8\%$ even at the closest settings.

\noindent \textbf{Power sweep.}
We fix the geometry at \SI{4.0}{m}, $0^{\circ}$, and \SI{1.5}{m}. The $0^{\circ}$ angle and camera-level source height provide direct, consistent illumination, while the \SI{4.0}{m} distance avoids the frequent sensor saturation observed at \SI{2.0}{m}. Together, these settings produce a strong but non-saturated response, allowing the effect of optical power to be isolated. We then vary optical power across \SIlist{2;4;8;12;20}{W}. Table~\ref{tab:sim_power_results} shows that nighttime attack strength increases monotonically with power, from $-15.08\%$ at \SI{2}{W} to $-80.20\%$ at \SI{20}{W}. Under daytime lighting, the same sweep produces only small changes, from $0.65\%$ to $3.48\%$.

\begin{table}[tbp]
\centering
\caption{Simulation power sweep results at \SI{4.0}{m} distance,
$0^{\circ}$ angle, and \SI{1.5}{m} height.} 
\label{tab:sim_power_results}

\footnotesize
\setlength{\tabcolsep}{3pt}
\renewcommand{\arraystretch}{1.05}

\begin{tabular}{@{}lccccc@{}}
\toprule
\textbf{Cond.} &
\textbf{2W} &
\textbf{4W} &
\textbf{8W} &
\textbf{12W} &
\textbf{20W} \\
\midrule

\rowcolor{gray!8}
Night &
$-15.1\%$ &
\cellcolor{red!14}\textbf{$-24.4\%$} &
\cellcolor{red!14}\textbf{$-40.3\%$} &
\cellcolor{red!14}\textbf{$-54.6\%$} &
\cellcolor{red!14}\textbf{$-80.2\%$} \\

Day &
$0.7\%$ &
$1.1\%$ &
$1.8\%$ &
$2.4\%$ &
$3.5\%$ \\

\bottomrule
\end{tabular}
\end{table}

\begin{figure}[t]
    \centering
    \includegraphics[width=0.47\columnwidth]{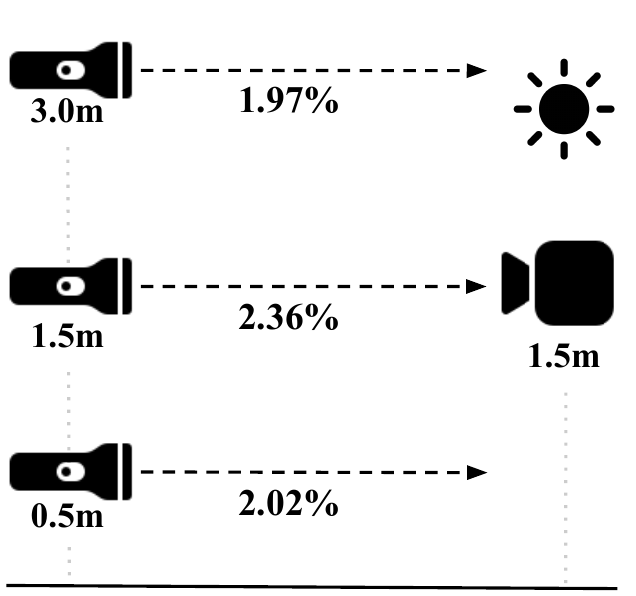}\hfill
    \includegraphics[width=0.47\columnwidth]{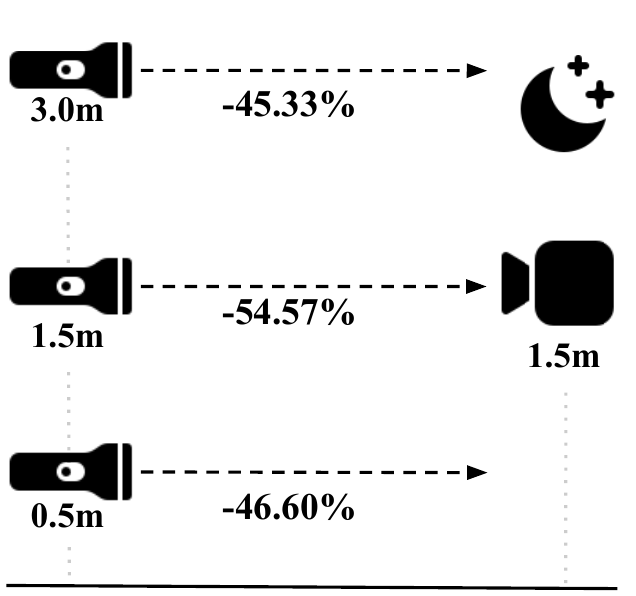}
    \caption{Simulation height sweep results at \SI{4.0}{m} distance, $0^{\circ}$ angle, and \SI{12}{W} flashlight power under daytime (left) and nighttime (right) ambient conditions.} 
    \label{fig:sim_height_results}
\end{figure}

\noindent \textbf{Height sweep.}
\noindent Using the same non-saturating distance of \SI{4.0}{m}, we fix the angle at $0^{\circ}$ and power at \SI{12}{W}, then vary source height across \SIlist{0.5;1.5;3.0}{m}. Figure~\ref{fig:sim_height_results} shows that camera-level placement at \SI{1.5}{m} gives the strongest response under both ambient conditions. At night, the score is $-54.57\%$ at \SI{1.5}{m}, compared with $-46.60\%$ at \SI{0.5}{m} and $-45.33\%$ at \SI{3.0}{m}. Daytime results follow the same pattern with smaller changes.

\begin{tcolorbox}[enhanced, colback=Cerulean!8, colframe=gray!70!black, toprule=0pt, bottomrule=0pt, rightrule=0pt, leftrule=2pt, boxsep=0pt, left=4pt, right=2pt, top=2pt, bottom=2pt]
\textbf{Simulation key findings.}
\ac{FLASH} is strongest under low ambient illumination, at short $d_{fc}$, near $\theta=0^\circ$, at camera-level $z_f$, and with high $P_f$. The spatial sweep reaches extreme darkening near \SI{2}{m} and $0^\circ$; degradation rises from $-15.08\%$ at \SI{2}{W} to $-80.20\%$ at \SI{20}{W} and peaks at $-54.57\%$ for $z_f=\SI{1.5}{m}$. Ambient illumination, distance, angle, and power dominate the response. 
\end{tcolorbox}

\subsection{Physical Camera Evaluation}
\label{subsec:physical_eval}
The physical camera evaluation addresses EQ2 through matched clean, continuous-light, randomized-frequency, and \ac{FLASH} conditions across black-box camera pipelines. We then organize the observed results by dominant response type. The parameter sweeps on the Raspberry Pi Camera Module 3 and iPhone 16 Pro further address EQ1 across distance, angle, source setting, measured ambient illuminance, and frame rate.

\subsubsection{Evaluation Design and Cross-Device Comparison}
\label{subsubsec:physical_cross_device}
\noindent \textbf{Black-box evaluation and controls.}
Because proprietary cameras do not expose RAW brackets or \ac{ISP} internals, we treat physical devices as black-box pipelines and evaluate only their final image or video output. Each camera is tested under four matched default conditions: (1) clean baseline, (2) continuous light (matched to the measured illuminance of the \ac{FLASH} source at the camera plane), (3) randomized-frequency flashing light, and (4) \ac{FLASH}. These controls separate timing-sensitive \ac{FLASH} behavior from ordinary bright-light exposure and randomized-frequency flashing.

\noindent \textbf{Default cross-device comparison.}
We first run the four matched conditions across all seven physical cameras using the default setup. The default setup uses a camera-to-source distance of \SI{2}{m}, attack angle of $0^{\circ}$, source height of \SI{1.5}{m}, low ambient illuminance, and the high flashlight setting. The Raspberry Pi HQ Camera was operated in its default single-exposure mode and serves as the non-HDR baseline. Its IMX477 sensor supports DOL-HDR, but this mode is not implemented by the stock Raspberry Pi camera stack. Raspberry Pi 5 separately provides an optional ISP-based computational HDR mode that accumulates frames captured with the same exposure setting. This mode differs from the temporal multi-exposure fusion targeted in this work and was not enabled during capture. We do not expect it to be immune to bright light because non-\ac{HDR} \ac{ISP} stages can still react to illumination changes. Instead, we use it to test whether the observed response is specific to \ac{HDR} or \ac{HDR}-relevant pipelines.

\noindent \textbf{Metric application and normalization.}
We use the signed luma-change, severe-darkening, extreme-darkening, and CNR metrics defined in Section~\ref{subsec:eval_questions_metrics} for the default cross-device physical-camera comparison. For the physical camera recordings, each non-clean condition is compared with a matched reference recording from the same camera and setup. The clean-normalized comparison uses the no-light baseline as the reference and is reported for completeness. To isolate timing-sensitive \ac{FLASH} behavior, our main default cross-device analysis uses control-normalized comparisons. In this analysis, \ac{FLASH} is compared against continuous light and randomized-frequency flashing under the same default setup. This separates timing-dependent degradation from ordinary bright-light exposure and randomized-frequency flashing. For the parameter sweeps, we use clean versus \ac{FLASH} pairs and report median signed decoded luma-unit change, because the goal of the sweep is to compare trends across distance, angle, source setting, ambient illuminance, and FPS rather than to summarize collapse rates.

\newcommand{\failcell}[1]{\cellcolor{red!14}\textbf{#1}}

\begin{table}[t]
\centering
\caption{Control-normalized full-frame response and dominant response under \ac{FLASH} in the default setup.} 
\label{tab:control_normalized_default}
\scriptsize
\setlength{\tabcolsep}{2pt}
\renewcommand{\arraystretch}{1.12}

\begin{tabularx}{\columnwidth}{@{}
>{\hsize=1.10\hsize\raggedright\arraybackslash}X
rrrr
>{\hsize=0.90\hsize\raggedright\arraybackslash}X
@{}}
\toprule
\textbf{Camera} &
\textbf{Frames} &
\textbf{Med. $\delta Y'$} &
\textbf{Sev.} &
\textbf{Ext.} &
\textbf{Dominant response} \\
&
&
\textbf{(\%)} &
\textbf{(\%)} &
\textbf{(\%)} & \\
\midrule

\rowcolor{gray!14}
\multicolumn{6}{@{}l@{}}{\textit{\textbf{\ac{FLASH} vs. continuous light}}} \\
\rowcolor{white}
iPhone 16 Pro & 1299 & 213.9 & \failcell{50.0} & \failcell{50.0} & Intermittent collapse \\
\rowcolor{gray!8}
Kodak PixPro FZ45 & 331 & 159.1 & 0.0 & 0.0 & Brightening/compensation \\
\rowcolor{white}
onn Indoor Camera & 292 & -4.6 & 0.0 & 0.0 & Localized degradation \\
\rowcolor{gray!8}
Raspberry Pi AI Camera & 143 & 350.9 & 0.0 & 0.0 & Brightening/compensation \\
\rowcolor{white}
Wyze Battery Cam Pro & 172 & -42.6 & \failcell{33.7} & \failcell{33.7} & Full-frame dark collapse \\
\rowcolor{gray!8}
Raspberry Pi Module 3 & 589 & 317.6 & 0.0 & 0.0 & Brightening/compensation \\
\rowcolor{white}
\textit{Raspberry Pi HQ (non-HDR)} & 300 & -41.5 & \failcell{27.0} & 1.0 & Mixed response \\

\midrule
\rowcolor{gray!14}
\multicolumn{6}{@{}l@{}}{\textit{\textbf{\ac{FLASH} vs. randomized flashing}}} \\
\rowcolor{white}
iPhone 16 Pro & 1299 & 10.4 & \failcell{25.1} & \failcell{24.9} & Intermittent collapse \\
\rowcolor{gray!8}
Kodak PixPro FZ45 & 351 & -6.8 & 0.0 & 0.0 & Mild darkening \\
\rowcolor{white}
onn Indoor Camera & 438 & -5.2 & 0.0 & 0.0 & Localized degradation \\
\rowcolor{gray!8}
Raspberry Pi AI Camera & 143 & 4.3 & 0.0 & 0.0 & Brightening/compensation \\
\rowcolor{white}
Wyze Battery Cam Pro & 155 & -46.8 & \failcell{33.5} & \failcell{33.5} & Full-frame dark collapse \\
\rowcolor{gray!8}
Raspberry Pi Module 3 & 589 & 162.3 & \failcell{19.7} & \failcell{8.0} & Intermittent collapse \\
\rowcolor{white}
\textit{Raspberry Pi HQ (non-HDR)} & 300 & 269.4 & \failcell{30.3} & 6.3 & Mixed response \\
\bottomrule
\end{tabularx}

\vspace{2pt}
\begin{minipage}{\columnwidth}
\scriptsize
\textit{Note:} Severe darkening denotes frames with $\delta Y' \leq -50\%$.
Extreme darkening denotes frames with $\delta Y' \leq -80\%$.
Highlighted cells indicate nonzero scene-wide severe-darkening or extreme-darkening failure rates.
\end{minipage}
\end{table}

\noindent \textbf{Cross-device results.}
Table~\ref{tab:control_normalized_default} shows that \ac{FLASH} produces pipeline-dependent responses rather than uniform darkening. The strongest full-frame collapse appears on the iPhone 16 Pro and Wyze Battery Cam Pro. The Raspberry Pi Camera Module 3 shows timing-sensitive degradation relative to randomized-frequency flashing, while the Kodak PixPro FZ45 and Raspberry Pi AI Camera primarily show brightening or exposure-compensation behavior. The onn Indoor Camera shows localized degradation without full-frame collapse. We therefore organize the results below by response type.

\subsubsection{Dark-Collapse and Timing-Sensitive Responses}
\label{subsubsec:collapse_responses}

\noindent \textbf{iPhone 16 Pro.}
The iPhone 16 Pro exhibits intermittent frame-level collapse. Relative to continuous light, \ac{FLASH} produces severe-darkening and extreme-darkening rates of 50.0\%. Relative to randomized-frequency flashing, the corresponding rates are 25.1\% and 24.9\%. This indicates intermittent frame-level collapse, even though the median signed luma change is positive in both comparisons.

\begin{figure}[t]
    \centering
    \includegraphics[width=\columnwidth]{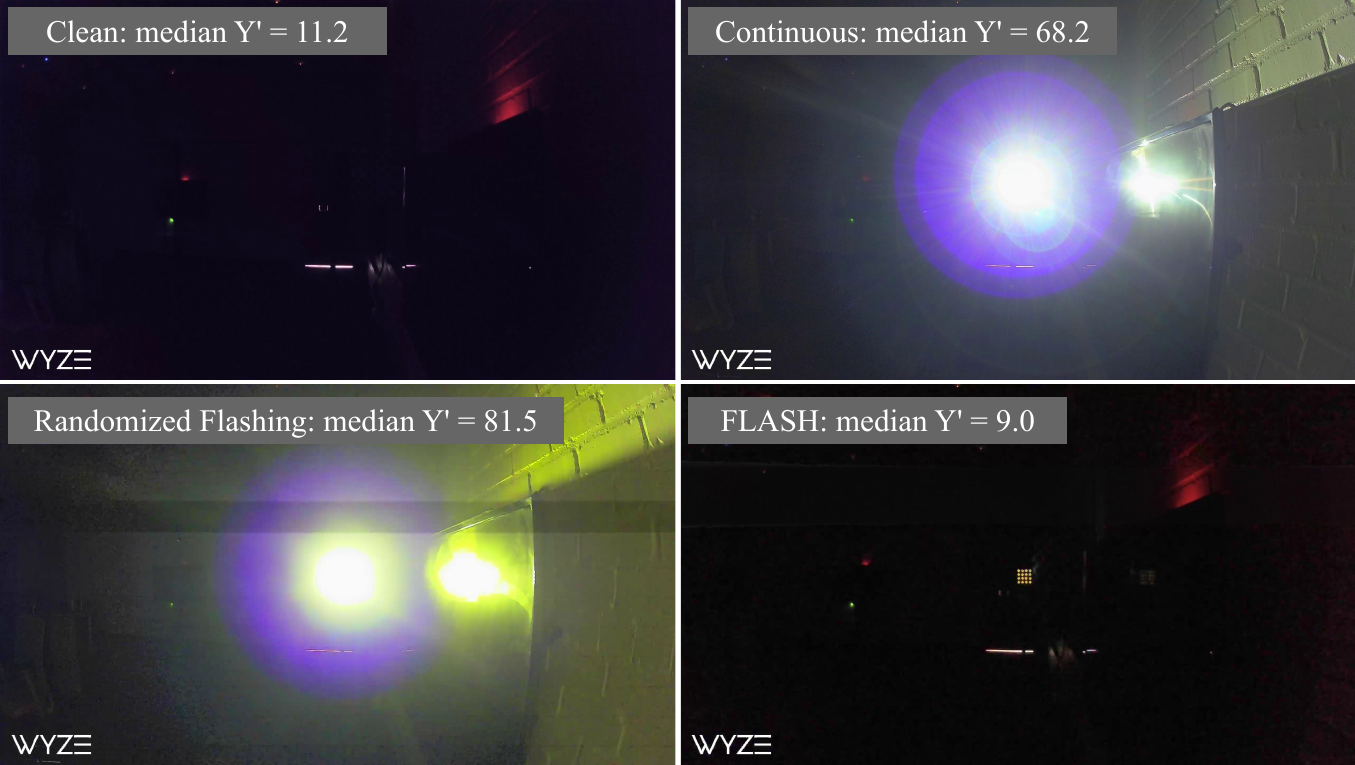}
    \caption{Four-condition Wyze Battery Cam Pro comparison.} Continuous light and randomized-frequency flashing raise median \(Y'\) from 11.2 to 68.2 and 81.5, while \ac{FLASH} reduces it to 9.0, which is darker than the clean scene (11.2), supporting timing-dependent collapse rather than bright-light interference.
    \label{fig:wyze_four_condition}
\end{figure}

\noindent \textbf{Wyze Battery Cam Pro.}
The Wyze Battery Cam Pro exhibits strong full-frame collapse. Relative to continuous light, \ac{FLASH} produces a median signed luma reduction of 42.6\%, with severe-darkening and extreme-darkening rates of 33.7\%. Relative to randomized-frequency flashing, the median signed luma reduction is 46.8\%, with severe-darkening and extreme-darkening rates of 33.5\%. Figure~\ref{fig:wyze_four_condition} shows that continuous and randomized-frequency flashing light increase the decoded median luma, while \ac{FLASH} reduces the output to a level close to the clean low-light baseline. This behavior indicates that the collapse is associated with pulse timing rather than bright illumination alone.

\noindent The Wyze camera also exhibits a built-in response that is not captured by frame-level luma metrics alone. Across 10 repeated \ac{FLASH} trials, all 10 trials activated the camera's automatic spotlight or infrared behavior. The same response occurred in 0 out of 10 continuous-light trials and 0 out of 10 randomized-frequency trials. We retain this automatic behavior because spotlight and infrared modes are part of normal surveillance-camera operation. This result indicates that the Wyze pipeline interpreted the \ac{FLASH}-affected scene as a low-visibility condition, while continuous and randomized-frequency flashes did not achieve the same treatment.

\noindent \textbf{Raspberry Pi Camera Module 3.}
The Raspberry Pi Camera Module 3 shows timing-sensitive degradation mainly relative to randomized-frequency flashing. In that comparison, 19.7\% of frames meet the severe-darkening threshold and 8.0\% meet the extreme-darkening threshold.

\noindent \textbf{Non-HDR interpretation boundary.}
The Raspberry Pi HQ Camera is treated as an interpretation boundary rather than as evidence of \ac{HDR}-specific failure. It is a non-\ac{HDR} camera, yet it shows severe-darkening rates of 27.0\% and 30.3\% in the two control-normalized comparisons. Its extreme-darkening rates remain much lower, at 1.0\% and 6.3\%, and its median response is inconsistent across controls: $-41.5\%$ relative to continuous light but $+269.4\%$ relative to randomized-frequency flashing. We therefore use it as evidence that strong optical stimulation can also disturb non-\ac{HDR} camera output, while the strongest extreme-darkening collapse appears on HDR or HDR-relevant black-box pipelines in our evaluated set.

\subsubsection{Brightening, Compensation, and Localized Responses}
\label{subsubsec:other_responses}

\noindent \textbf{Raspberry Pi AI Camera.}
On the Raspberry Pi AI Camera, \ac{FLASH} produces a brightening response. The full-frame median change is $+350.9\%$ relative to continuous light and $+4.3\%$ relative to randomized-frequency flashing. Its target-region median changes are $+398.1\%$ and $+9.4\%$ under the same two comparisons. These results indicate exposure-compensation behavior rather than full-frame collapse.

\noindent \textbf{Kodak PixPro FZ45.}
On the Kodak PixPro FZ45, \ac{FLASH} is much brighter than continuous light in the full-frame metric, with a median $\delta Y'$ of $+159.1\%$, and it also increases target-region luma by $+185.6\%$ relative to continuous light. Relative to randomized-frequency flashing, Kodak shows a small full-frame reduction of $-6.8\%$ and a stronger background-region reduction of $-26.1\%$, although these changes do not reach the severe-darkening threshold.

\noindent \textbf{onn Indoor Camera.}
The onn Indoor Camera's regional measurements show stronger localized degradation. Relative to continuous light, the gray-card region decreases by 38.8\%, the target region decreases by 22.0\%, and target-background CNR decreases by 48.8\%. This ISP does not exhibit full-frame extreme-darkening under the tested conditions, although its median full-frame luma change is \(-4.6\%\) relative to continuous light and \(-5.2\%\) relative to randomized-frequency flashing. The onn results therefore indicate localized visibility and contrast loss rather than full-frame collapse.

\begin{tcolorbox}[enhanced, colback=Cerulean!8, colframe=gray!70!black, toprule=0pt, bottomrule=0pt, rightrule=0pt, leftrule=2pt, boxsep=0pt, left=4pt, right=2pt, top=2pt, bottom=2pt]
\textbf{Physical camera key findings.}
\ac{FLASH} causes pipeline-dependent responses. Extreme-darkening reaches 50.0\%/24.9\% on the iPhone and 33.7\%/33.5\% on the Wyze camera relative to continuous/randomized flashing. Wyze triggers its low-visibility response in 10/10 \ac{FLASH} trials versus 0/10 controls. Other devices show collapse, brightening, compensation, or localized degradation shaped by geometry, ambient illumination, and capture mode.
\end{tcolorbox}

\subsubsection{Operational Robustness Evaluation}
\label{subsubsec:physical_robustness}

To further address EQ1, we perform matched clean-versus-\ac{FLASH} physical parameter sweeps on the iPhone 16 Pro and Raspberry Pi Camera Module 3, the two evaluated devices supporting selectable \ac{fps} modes. We vary camera-to-source distance \(d_{fc}\), aiming angle \(\theta\), source setting \(P_f\), measured ambient illuminance \(A\), and camera frame rate \(f_{\mathrm{fps}}\). We report median signed decoded luma-unit change \(\Delta Y'\) because percent-normalized changes become unstable when low-light clean baselines approach zero. Table~\ref{tab:operational_robustness_summary} summarizes the results; the dataset composition, complete sweep plot, and per-setting analysis are provided in Appendix~\ref{app:physical_robustness}.

\begin{table*}[t]
    \centering
    \caption{Physical operational robustness under matched clean-versus-FLASH trials. Entries report median signed decoded luma-unit change \(\Delta Y'\) for the settings listed in order.}
    \label{tab:operational_robustness_summary}
    \small
    \setlength{\tabcolsep}{4pt}
    \renewcommand{\arraystretch}{1.12}

    \begin{tabularx}{\textwidth}{@{}
        >{\raggedright\arraybackslash}p{2.9cm}
        >{\raggedright\arraybackslash}p{2.4cm}
        >{\raggedright\arraybackslash}p{3.6cm}
        >{\raggedright\arraybackslash}p{3.1cm}
        >{\raggedright\arraybackslash}X
        @{}}
        \toprule
        \textbf{Parameter} &
        \textbf{Settings} &
        \textbf{iPhone 16 Pro} &
        \textbf{Module 3} &
        \textbf{Trend} \\
        \midrule

        \rowcolor{gray!8}
        Distance \(d_{fc}\) &
        \(2, 4, 6, 8~\mathrm{m}\) &
        \(60.60, 28.85, 16.90, 12.62\) &
        \(2.82, 0.74, 0.29, 0.42\) &
        Decreases with distance. \\

        \rowcolor{white}
        Aiming angle \(\theta\) &
        \(0^{\circ}, 10^{\circ}, 20^{\circ}, 30^{\circ}\) &
        \(60.60, 41.15, 17.53, 14.66\) &
        \(2.82, 1.92, 0.46, 0.50\) &
        Strongest near axis. \\

        \rowcolor{gray!8}
        Source setting \(P_f\) &
        \(25\%, 50\%, 100\%\) &
        \(39.19, 52.59, 32.94\) &
        \(2.18, 0.71, 2.60\) &
        Non-monotonic. \\

        \rowcolor{white}
        Ambient illuminance \(A\) &
        \(0, 27.3, 239~\mathrm{lux}\) &
        \(60.44, -6.12, -4.51\) &
        \(5.98, 1.00, -0.12\) &
        Suppressed by ambient illuminance. \\

        \rowcolor{gray!8}
        Frame rate \(f_{\mathrm{fps}}\) &
        \(24, 30, 60, 120~\mathrm{fps}\) &
        \(67.39, 65.87, 45.43, 52.61\) &
        N/A, \(12.03, 6.56, 11.83\) &
        Capture-mode dependent. \\

        \bottomrule
    \end{tabularx}
\end{table*}

Distance and off-axis aiming attenuate the response on both cameras, while increased ambient illuminance strongly suppresses the low-light response. Source setting and frame rate produce non-monotonic, device-dependent results, consistent with proprietary exposure, tone-mapping, and capture-mode processing. Although response magnitude varies across capture modes, \(\Delta Y'\) remains nonzero at every tested frame rate.

\subsection{System Case Study: comma 3X Camera-Stream Degradation and OpenPilot Path State}
\label{subsec:system_eval}

The system case study addresses EQ3 through controlled stationary comma 3X/OpenPilot trials. The vehicle remains parked throughout each trial. We evaluate three conditions: an unobstructed baseline with no target or attack equipment, a target-present control without \ac{FLASH}, and a target-present \ac{FLASH} condition. During the \ac{FLASH} condition, the source is \SI{3}{m} directly in front of the forward-facing camera at \(0^\circ\) and a height of \SI{1.5}{m}. This favorable stress-testing geometry is selected based on the preceding simulation and component-level evaluations. The evaluated targets are a speed-limit sign, an orange traffic cone, and a pedestrian, together with general road-scene visibility.

\noindent \textbf{Displayed path state.}
For each condition, we record whether OpenPilot displays a green planned path, a red blocked path, or no path. This is an interface-level observation; we do not observe OpenPilot's internal perception or planning state.

\noindent \textbf{Offline camera-stream processing.}
Separately, we export the comma 3X camera streams and compare each \ac{FLASH} video with its corresponding no-\ac{FLASH} reference. Target-present trials use the target-present control as the reference, while the general road-scene trial uses the unobstructed baseline. We compute full-frame signed luma change, target-region signed luma change, target-background signed CNR change, severe-darkening rate, extreme-darkening rate, and representative reference-versus-\ac{FLASH} frames. The traffic-sign and cone targets are analyzed from the same video pair, so their full-frame metrics are reported once. Their target-region and CNR metrics are reported separately because they use different object ROIs. Table~\ref{tab:system_level_comma3x} summarizes these measurements.

\newcommand{\extcell}[1]{\cellcolor{red!14}\textbf{#1}}

\begin{table*}[h]
\centering
\caption{Offline comma 3X camera-stream degradation under \ac{FLASH} relative to the corresponding no-\ac{FLASH} references. Full-frame metrics capture scene-level effects, while target ROI and CNR metrics capture object-level visibility. Highlighted cells indicate the strongest observed effects.}

\label{tab:system_level_comma3x}

\setlength{\tabcolsep}{2pt}
\renewcommand{\arraystretch}{1.08}

\resizebox{0.8\textwidth}{!}{%
\begin{tabular}{@{}l l c c c c c@{}}
\toprule
\textbf{Target/Condition} &
\textbf{Meas.} &
\textbf{Frames} &
\textbf{Med. $\delta Y'$ (\%)} &
\textbf{Min. $\delta Y'$ (\%)} &
\textbf{Sev. (\%)} &
\textbf{Min. $\delta\mathrm{CNR}$ (\%)} \\
\midrule

\rowcolor{gray!8}
Shared sign/cone scene & Full frame & 100 & 86.6 & -38.7 & 0.0 & -- \\

\rowcolor{white}
Speed-limit sign & Target ROI & 100 & 101.9 & 21.4 & 0.0 & 44.2 \\

\rowcolor{gray!8}
Cone & Target ROI & 100 & 234.2 & \extcell{-61.8} & \extcell{23.0} & \extcell{-90.8} \\

\rowcolor{white}
Pedestrian & Target ROI & 560 & \extcell{305.3} & \extcell{204.4} & 0.0 & \extcell{210.6} \\

\rowcolor{gray!8}
Road-scene visibility & Full frame & 120--560 & 63.5--73.0 & 56.6--65.2 & 0.0 & -- \\

\bottomrule
\end{tabular}%
}
\end{table*}

\noindent \textit{OpenPilot Displayed Path State.}
\label{subsubsec:path_availability_eval}
In the unobstructed baseline, OpenPilot displays a green planned path. When an object of interest is present without \ac{FLASH}, OpenPilot displays a red blocked path. Under \ac{FLASH}, it displays neither the green planned path nor the red blocked path. The missing displayed path is therefore distinct from the red blocked-path state observed in the target-present control.

\begin{figure*}[t]
    \centering
    \begin{subfigure}[t]{0.49\textwidth}
        \centering
        \includegraphics[width=\linewidth]{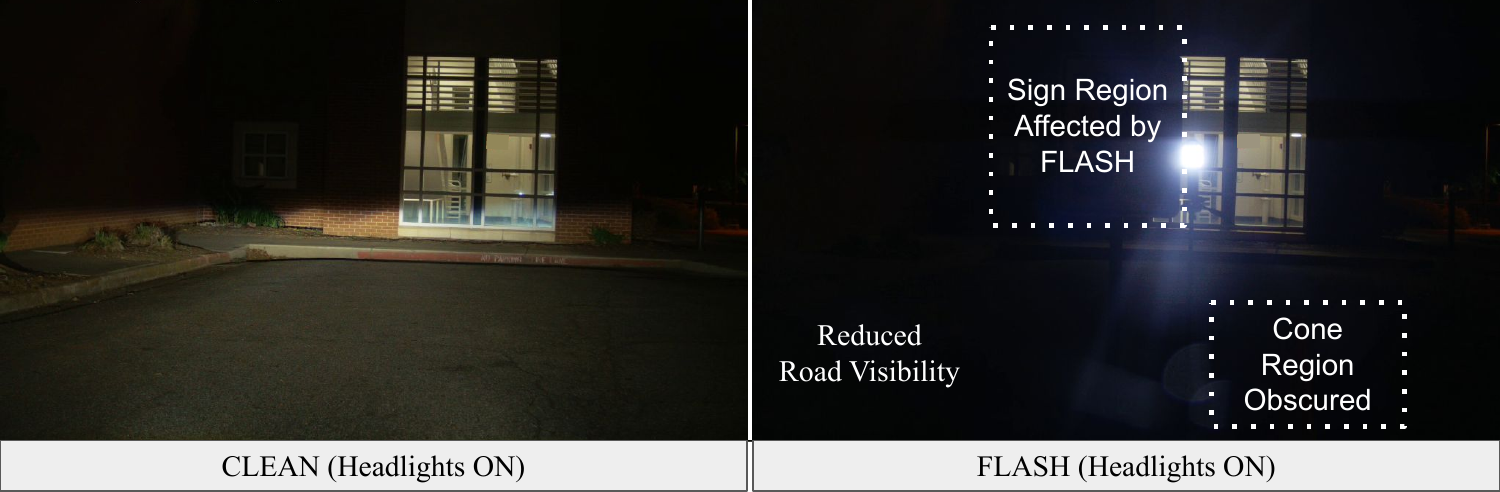}
        \caption{Traffic-sign and cone trial.}
        \label{fig:openpilot_signcone}
    \end{subfigure}
    \hfill
    \begin{subfigure}[t]{0.49\textwidth}
        \centering
        \includegraphics[width=\linewidth]{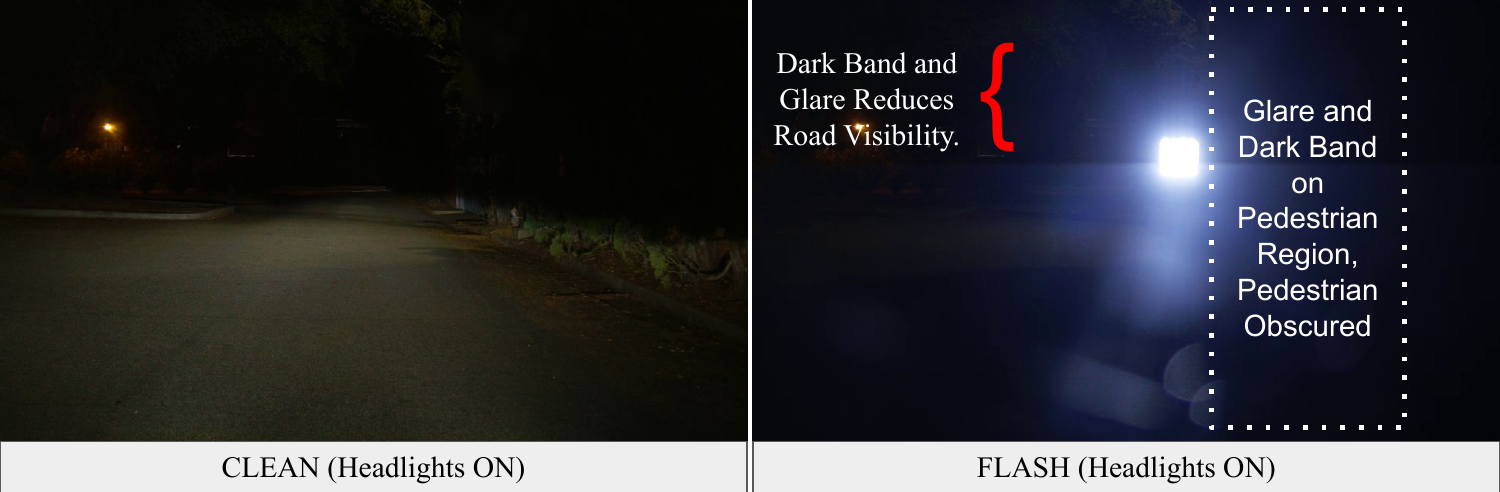}
        \caption{Pedestrian trial.}
        \label{fig:openpilot_ped}
    \end{subfigure}
    \caption{Representative comma 3X unobstructed baseline reference and \ac{FLASH} frames. Dashed boxes mark the evaluated target regions. In the sign and cone trial, \ac{FLASH} reduces traffic-sign visibility and obscures the cone region. In the pedestrian trial, \ac{FLASH} introduces glare and a dark horizontal band that reduces local road visibility. Quantitative results summarized in Table~\ref{tab:system_level_comma3x}.}
    \label{fig:openpilot_system_examples}
\end{figure*}

\noindent \textit{Cone Visibility.}
\label{subsubsec:obstacle_eval}
The cone target provides the strongest quantitative system-level degradation. Although the shared full-frame sign/cone scene does not reach the severe-darkening threshold, the cone ROI shows intermittent target-level collapse. The cone target ROI reaches a minimum $\delta Y'$ of $-61.8\%$, and 23.0\% of compared frames meet the severe-darkening threshold. The target-background CNR also shows strong intermittent loss, with a worst-frame \(\delta\mathrm{CNR}\) of $-90.8\%$ and 23.0\% of frames below the $-80\%$ CNR-change threshold. This indicates that \ac{FLASH} can substantially reduce the visibility and separability of a road object in the exported comma 3X stream, even when the full-frame average does not collapse. Figure~\ref{fig:openpilot_signcone} shows the representative clean-versus-\ac{FLASH} sign and cone example used for this system-level visualization.

\begin{figure}[t]
    \centering
    \includegraphics[width=0.75\columnwidth]{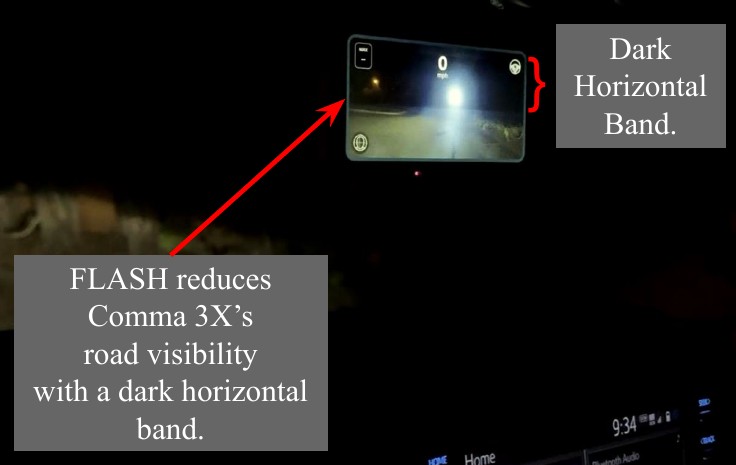} 
    \caption{In-car OpenPilot view during the controlled evaluation. Under \ac{FLASH}, the comma 3X stream contains a dark horizontal band that reduces local road visibility, and the interface displays neither a green planned path nor a red blocked path. Quantitative camera-stream results reported in Table~\ref{tab:system_level_comma3x}.}
    \label{fig:comma3x_incar}
\end{figure}

\noindent \textit{Road-Scene Visibility.}
\label{subsubsec:scene_visibility_eval}
The general road-scene trials show that \ac{FLASH} introduces visible glare, flare, and local visibility obstruction in the comma 3X stream. Figure~\ref{fig:comma3x_incar} shows the corresponding in-car OpenPilot view during the controlled evaluation, where the displayed comma 3X camera stream contains a dark horizontal band under \ac{FLASH}. Across three \ac{FLASH} trials, full-frame median $\delta Y'$ remains positive, ranging from $+63.5\%$ to $+73.0\%$, because the attack source adds substantial light to the scene. The worst full-frame values also remain positive, ranging from $+56.6\%$ to $+65.2\%$.

\noindent \textit{Pedestrian Visibility.}
\label{subsubsec:pedestrian_eval}
The pedestrian trial evaluates \ac{FLASH}-induced artifacts around a person in the deployed comma 3X front-camera stream. Here, the dominant effect is not target-region darkening but glare and local saturation near the injected light source. The pedestrian ROI becomes brighter under the luma metric, with a median \(\delta Y'\) of \(+305.3\%\), a minimum \(\delta Y'\) of \(+204.4\%\), and 0.0\% severe-darkening or extreme-darkening frames. The target-background CNR also increases, with a median \(\delta\mathrm{CNR}\) of \(+313.0\%\). We therefore report this trial as a glare and saturation artifact case rather than an extreme-darkening failure case. Figure~\ref{fig:openpilot_ped} shows the representative pedestrian example, where \ac{FLASH} introduces glare and a dark horizontal band that reduces local road visibility.

\noindent \textit{Traffic Sign Visibility.}
\label{subsubsec:traffic_sign_eval}
The traffic-sign trial evaluates a speed-limit sign in the exported comma 3X front-camera stream. In this case, the dominant effect is not target-region darkening but optical interference in the surrounding camera stream. The representative frames show glare, lens flare, and localized visibility obstruction near the injected light source. The selected speed-limit sign ROI becomes brighter under the luma metric, with a median \(\delta Y'\) of \(+101.9\%\), a minimum \(\delta Y'\) of \(+21.4\%\), and 0.0\% severe-darkening or extreme-darkening frames. We therefore treat this trial as a scene-visibility artifact case rather than an extreme-darkening failure case. The traffic-sign portion of Figure~\ref{fig:openpilot_signcone} shows this effect qualitatively, with reduced sign-region visibility under \ac{FLASH}.


\begin{tcolorbox}[enhanced, colback=Cerulean!8, colframe=gray!70!black, toprule=0pt, bottomrule=0pt, rightrule=0pt, leftrule=2pt, boxsep=0pt, left=4pt, right=2pt, top=2pt, bottom=2pt]
\textbf{System-level key findings.}
OpenPilot displays a green path in the unobstructed baseline and a red blocked path in the target-present control, but neither under \ac{FLASH}. Offline analysis shows scene-dependent artifacts, including 23.0\% severe cone-region darkening and a worst-case CNR loss of 90.8\%. The sign and pedestrian trials instead show glare, lens flare, and local visibility obstruction.
\end{tcolorbox}

%% file: 5-Evaluation_Setup.tex
\noindent\textbf{Simulation Environment.}
We implement the simulation in Blender within a controlled indoor environment under daytime and nighttime illumination baselines. The evaluated attack-source power \(P_f\) ranges from \SI{2}{W} to \SI{20}{W}. Complete scene-construction details and the representative source classes associated with these power levels are provided in Appendix~\ref{app:sim_imp_details}.

\noindent\textbf{Component-Level Hardware Setup.}
We evaluate seven physical cameras as black-box imaging pipelines in a fixed indoor setup. For the default trials, the camera and flash source are mounted at \(z_c=z_f=\SI{1.5}{m}\), the camera-to-source distance is \(d_{fc}=\SI{2}{m}\), and the target board is placed \SI{2}{m} from the camera. The manufacturer-rated \(P_f=\SI{2}{W}\) source is aimed toward the camera and target region. Ambient condition \(A\) and control-light illuminance are measured at the camera plane, and floor markings maintain consistent geometry across devices. We record the final image or video output using the available or default capture mode. The physical default uses \SI{2}{W} at \SI{2}{m}, while the simulation uses \SI{12}{W} at \SI{4}{m} to isolate parameter effects without trivially saturating the closest condition. Table~\ref{tab:physical_platforms} summarizes the evaluated platforms and their roles. Complete setup details and a photograph are provided in Appendix~\ref{app:physical_setup}.

\newcommand{\camicon}[1]{\raisebox{-0.25em}{\includegraphics[height=3.5em]{#1}}}
\newcommand{\tightcell}[1]{\begin{tabular}[c]{@{}c@{}}#1\end{tabular}}
\newcolumntype{C}[1]{>{\centering\arraybackslash}m{#1}}

\begin{table*}[tbp]
    \centering
    \caption{Physical platforms and their roles in the \ac{FLASH} evaluation. The first seven platforms are evaluated at the component physical level, while comma 3X is used for the system-level OpenPilot case study.}
    \label{tab:physical_platforms}
    \scriptsize
    \setlength{\tabcolsep}{1pt}
    \renewcommand{\arraystretch}{1.15}

    \begin{tabular}{@{}C{0.095\textwidth}*{8}{C{0.108\textwidth}}@{}}
        \toprule

        \rowcolor{white}
        & \camicon{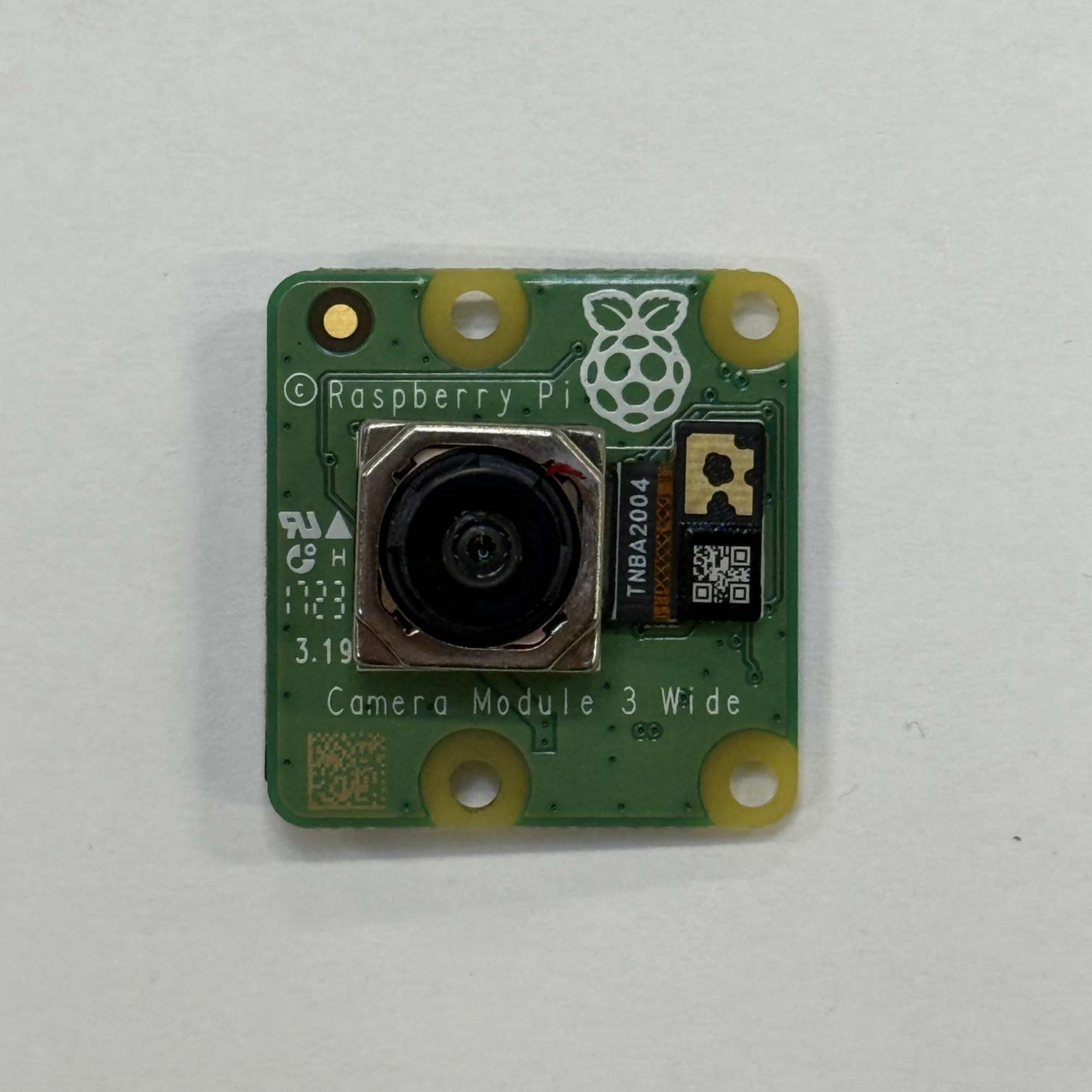}
        & \camicon{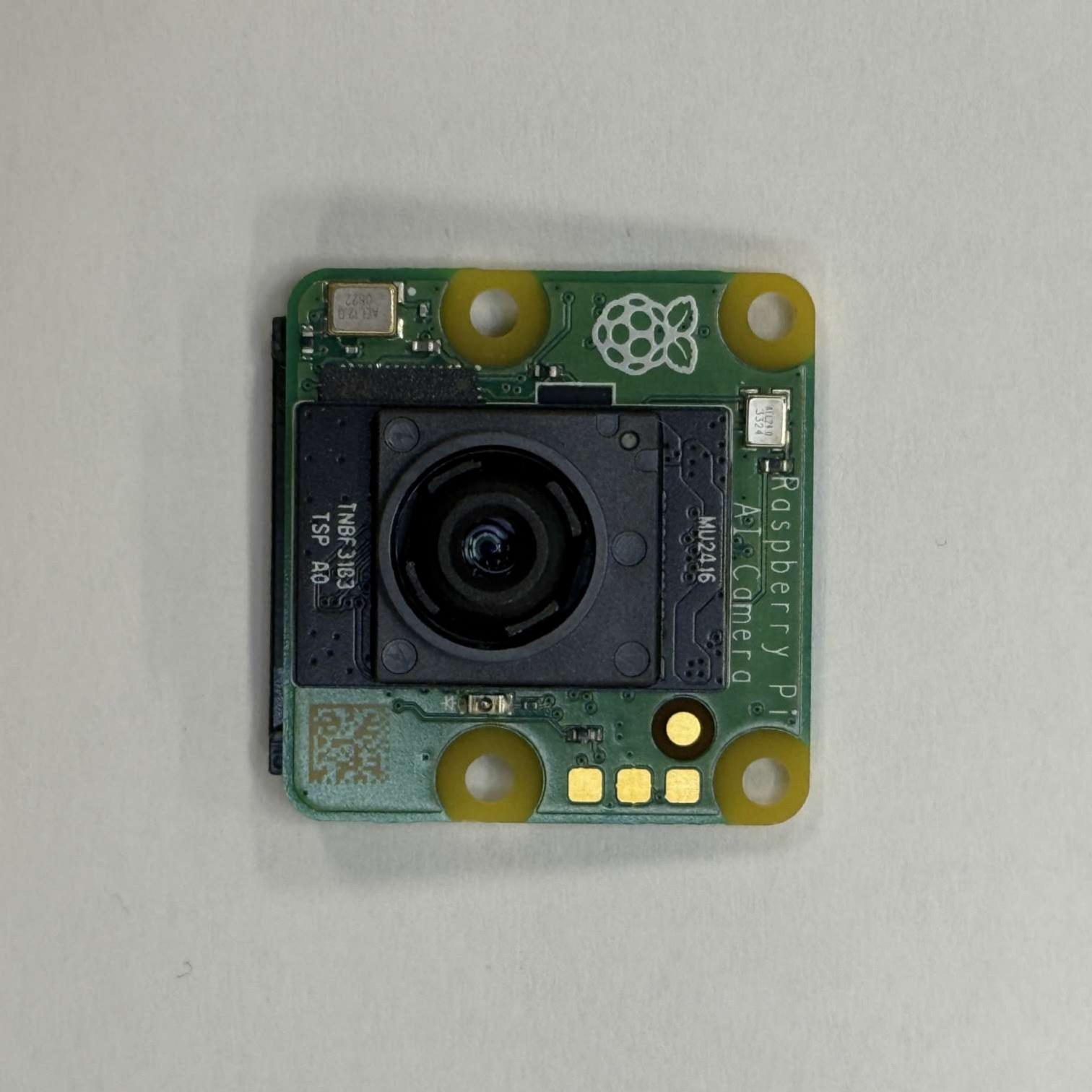}
        & \camicon{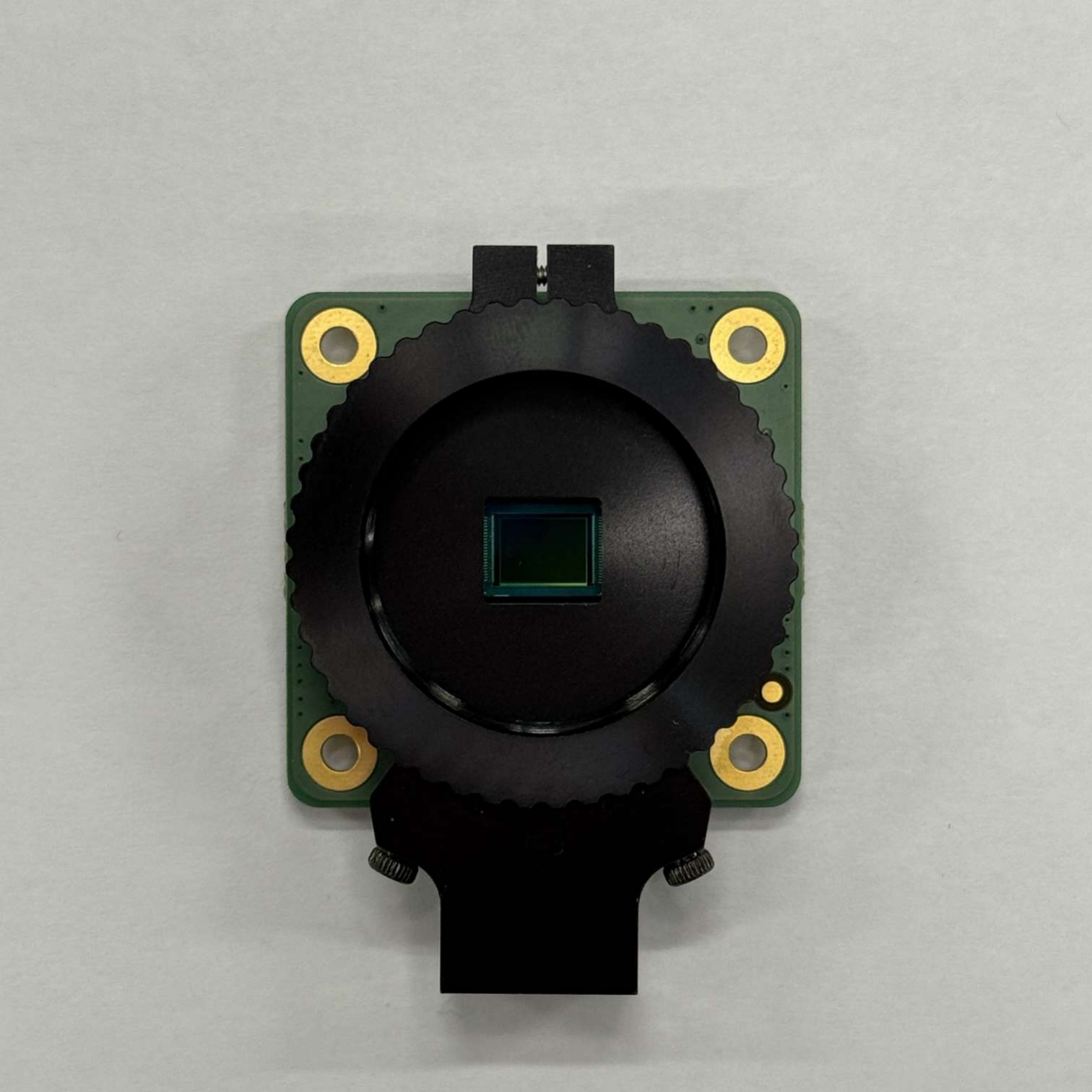}
        & \camicon{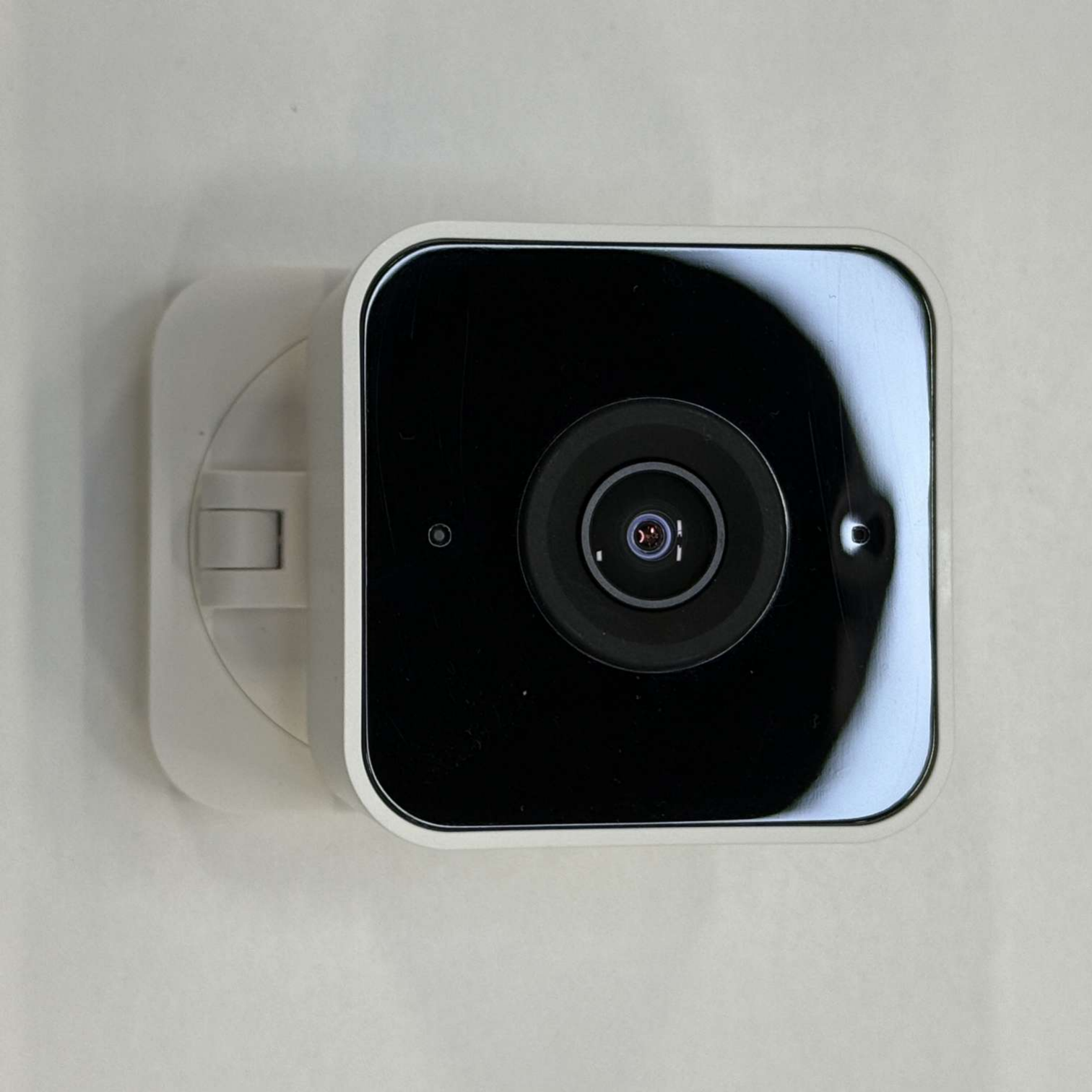}
        & \camicon{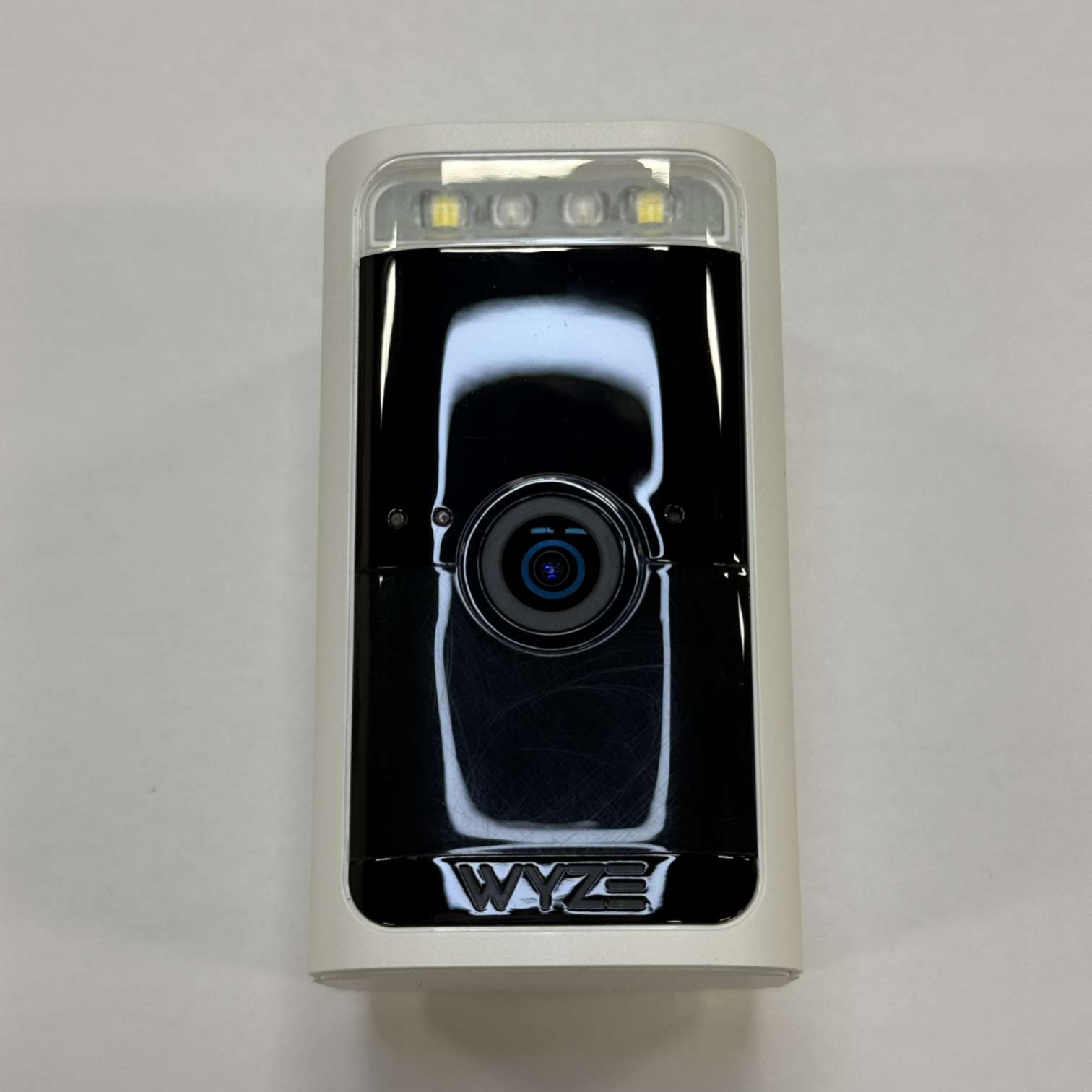}
        & \camicon{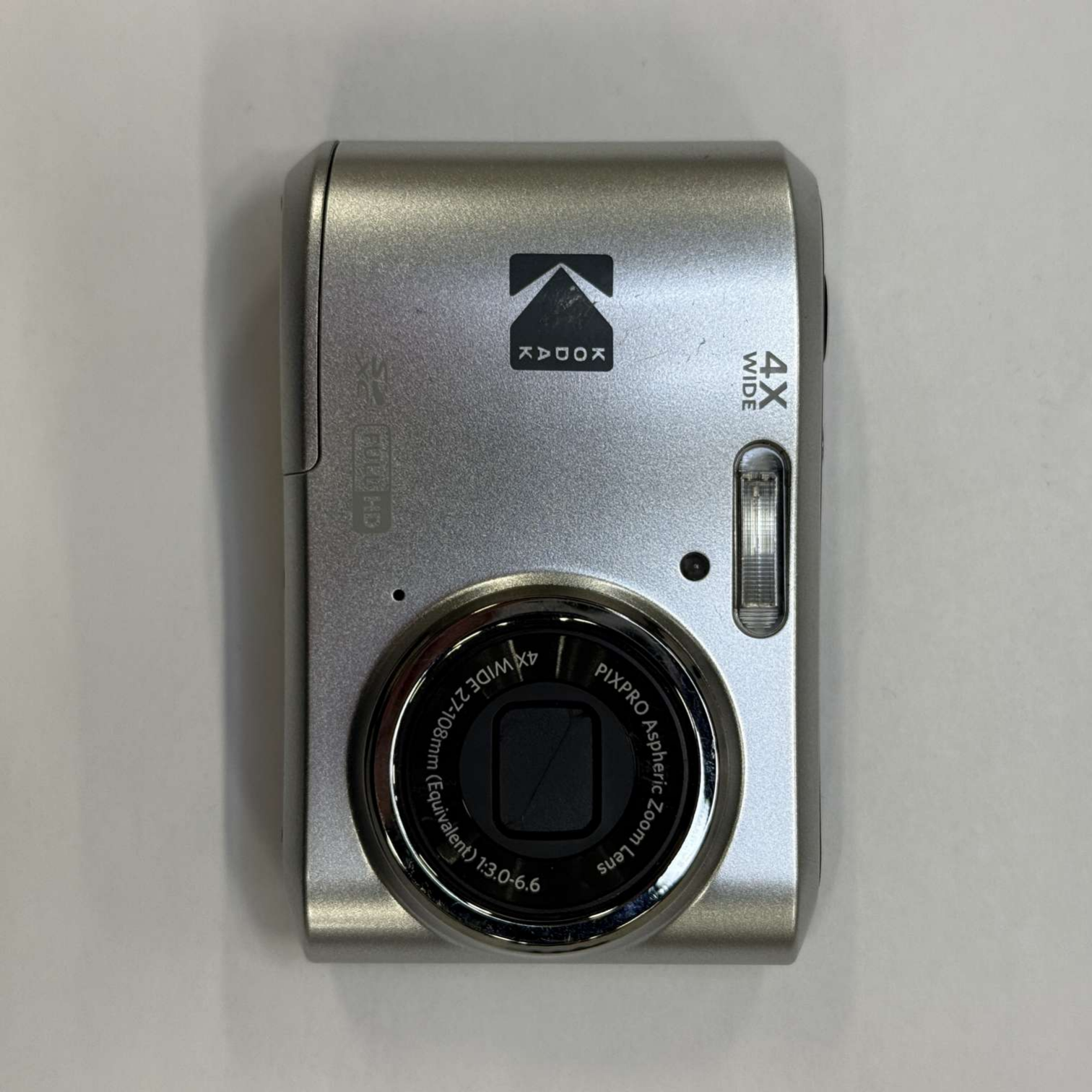}
        & \camicon{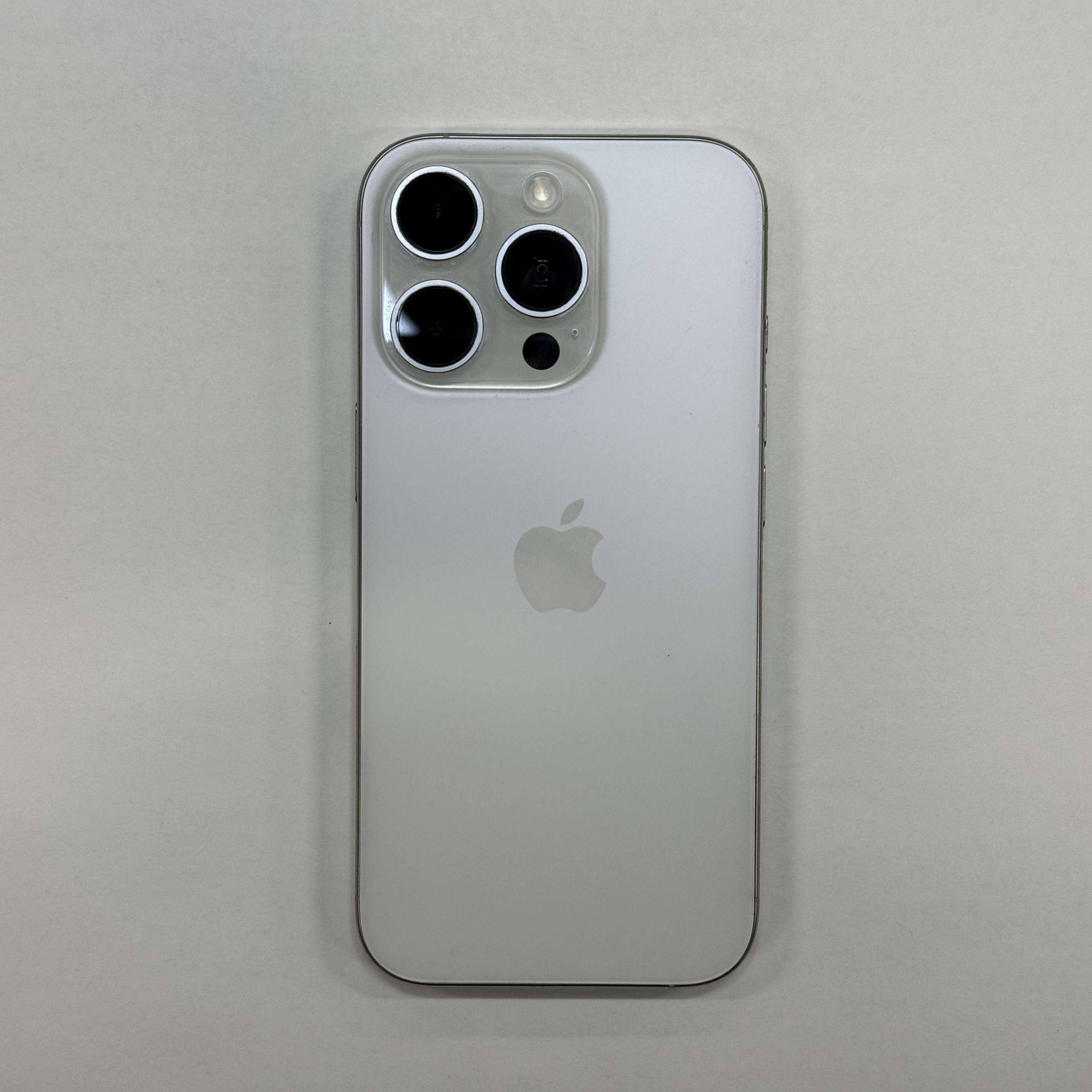}
        & \camicon{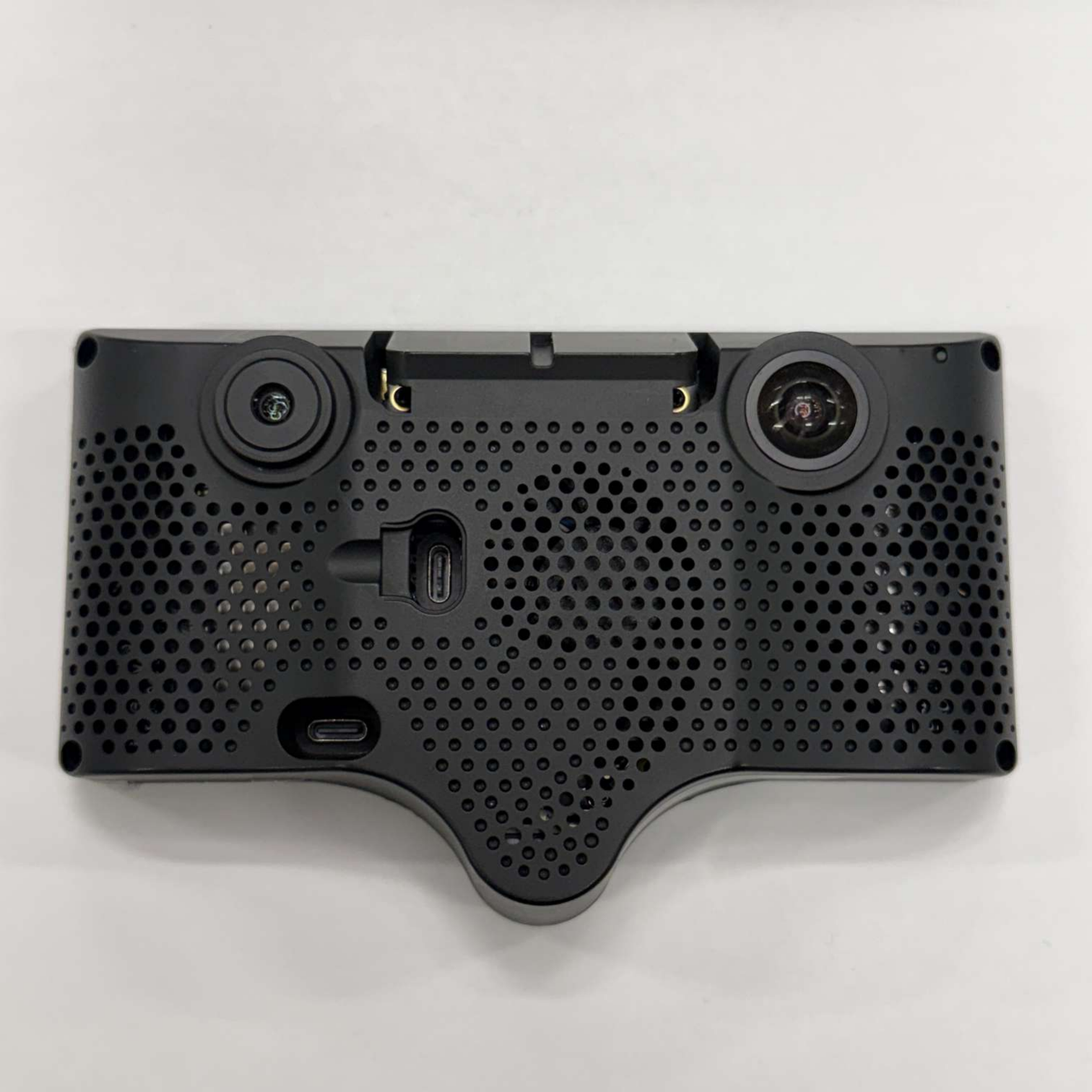} \\

        \midrule

        \rowcolor{gray!8}
        \textbf{Device}
        & \tightcell{Raspberry Pi\\Camera Module 3}
        & \tightcell{Raspberry Pi\\AI Camera}
        & \tightcell{Raspberry Pi\\HQ Camera}
        & \tightcell{onn\\Indoor Camera}
        & \tightcell{Wyze Battery\\Cam Pro}
        & \tightcell{Kodak PixPro\\FZ45}
        & \tightcell{iPhone 16 Pro}
        & \tightcell{comma 3X} \\

        \rowcolor{white}
        \textbf{Platform Class}
        & Embedded
        & Embedded
        & Embedded
        & Surveillance
        & Surveillance
        & Consumer
        & Smartphone
        & AV system \\

        \rowcolor{gray!8}
        \tightcell{\textbf{Evaluation Use}}
        & \tightcell{Mechanism tests;\\parameter sweeps}
        & \tightcell{Cross-device\\comparison}
        & \tightcell{Non-HDR\\baseline}
        & \tightcell{Surveillance\\camera\\evaluation}
        & \tightcell{Surveillance\\camera\\evaluation}
        & \tightcell{Photography\\evaluation}
        & \tightcell{Smartphone\\evaluation;\\parameter sweeps}
        & \tightcell{Controlled outdoor\\case study} \\

        \bottomrule
    \end{tabular}
\end{table*}

\noindent\textbf{System-Level comma 3X OpenPilot Setup.}
For the system-level case study, the comma 3X operates through its normal camera and OpenPilot pipeline in a controlled outdoor setting while the vehicle remains stationary. Matched target-present control and \ac{FLASH} trials keep the experimenter, source, target, and vehicle in the same positions; the control keeps the source inactive, while the \ac{FLASH} condition differs only in the emitted pulses. The source uses \(P_f=\SI{2}{W}\) and is placed at \(d_{fc}=\SI{3}{m}\), \(\theta=0^{\circ}\), and \(z_f=\SI{1.5}{m}\). We evaluate a speed-limit sign, traffic cone, pedestrian, and general road scene. For each trial, we record whether OpenPilot displays a green planned path, a red blocked-path indication, or no path, and we analyze full-frame and target-region degradation from the exported camera stream.

%% file: 7-Defense.tex
\begin{figure}[h]
    \centering
    \includegraphics[width=\columnwidth]{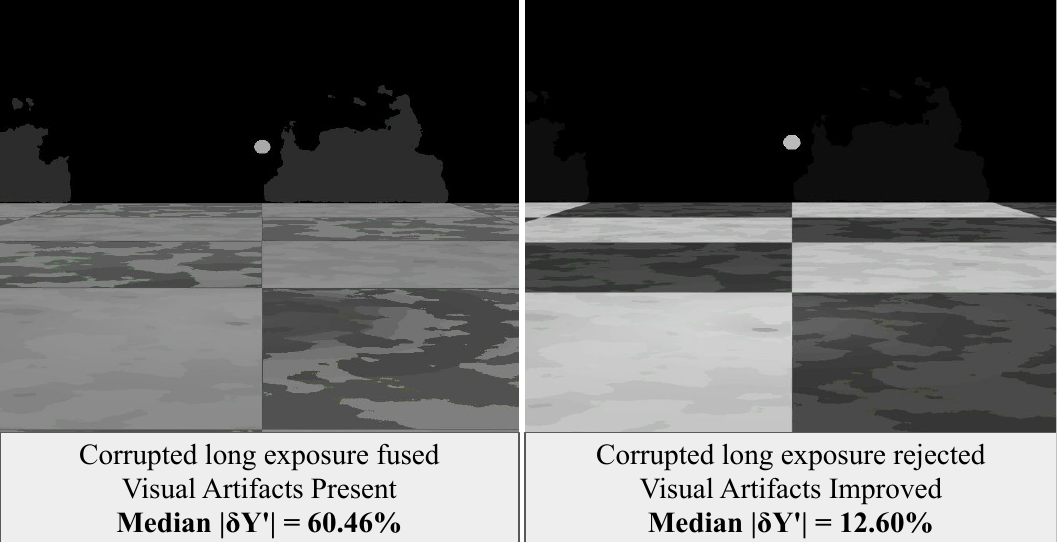}
    
    \vspace{0.2em}
    \makebox[0.49\columnwidth][c]{\small (a) Undefended \acs{FLASH}}%
    \hfill
    \makebox[0.49\columnwidth][c]{\small (b) Defended against \acs{FLASH}}
    
    \caption{Exposure-level rejection in the controlled night-only \ac{HDR} stress test. Rejecting the corrupted long exposure before fusion succeeds in all 756 cases with a 0.0\% clean false rejection rate and reduces median absolute output-luma deviation from 60.46\% to 12.60\% (79.16\%).}
    \label{fig:defense_comparison}
\end{figure}

\ac{FLASH} creates cross-bracket inconsistency by causing one or more exposures in a temporal \ac{HDR} sequence to receive illumination that is inconsistent with the rest of the bracket. Because this inconsistency is before exposure fusion, mitigation can act on the exposure evidence before the final frame is formed. We evaluate a proof-of-concept exposure-level rejection method in a controlled three-exposure setting in which corruption is injected into the long exposure. This allows us to test whether removing anomalous exposure evidence before fusion can reduce the resulting reconstruction deviation.

To support controlled analysis of internal \ac{HDR} behavior, we also developed \ac{HDR}-Lab, a modular open-source computational photography framework. Unlike the proprietary black-box camera pipelines used in the physical evaluation, \ac{HDR}-Lab exposes intermediate stages and allows controlled manipulation of individual exposure inputs. Implementation details and the access link are provided in Appendix~\ref{app:hdrlab}.

\noindent \textbf{Temporal exposure rejection.}
We propose an exposure-level anomaly filter that operates inside the \ac{HDR} reconstruction pipeline before exposure fusion. Conceptually, the filter tests the exposures in a bracket for lighting inconsistencies. In our evaluated implementation, each bracket contains short, medium, and long exposures. For each exposure, the filter computes a luminance statistic and compares it with the corresponding clean or recent reference behavior. If one exposure produces an abnormal luminance deviation, the filter flags that exposure as corrupted.

A flagged exposure is excluded from fusion or assigned a near-zero fusion weight, and the output frame is reconstructed from the remaining exposures. In our three-exposure implementation, the defense rejects at most one corrupted exposure and preserves the rest of the bracket. This differs from whole-frame rejection because the output frame can still be reconstructed from the remaining exposure evidence.

The defense targets the cross-bracket inconsistency exploited by \ac{FLASH}. If an anomalous exposure remains in the bracket, it can influence reconstruction before subsequent \ac{ISP} stages render the final frame. Exposure rejection instead removes the detected inconsistent exposure before fusion. Our evaluation tests this principle specifically for corruption of the long exposure in a three-exposure bracket.

\noindent \textbf{Controlled defense evaluation.}
Commercial cameras do not expose their internal exposure buffers, so we evaluate the defense in a controllable \ac{HDR} reconstruction pipeline rather than inside proprietary on-chip \ac{ISP} firmware. We use the night EXR simulation scenes as radiance inputs and construct synthetic three-exposure \ac{HDR} brackets containing short, medium, and long exposures. The stress test injects corruption only into the long exposure. Then the defense system applies exposure-level rejection before fusion, and we compare the reconstructed \ac{FLASH} output before and after defense.

Across 756 night-only stress-test cases, the detector rejects the corrupted long exposure in 756/756 cases, with a 0.0\% clean false rejection rate. The defense reduces the median absolute luminance deviation from 60.46\% before defense to 12.60\% after defense, a 47.86 percentage-point reduction and a 79.16\% relative reduction. The unoptimized Python/OpenCV prototype adds 108.85\,ms median latency per \ac{HDR} bracket, so this timing result is implementation-specific. This evaluation does not claim modification of closed commercial \ac{ISP} implementations or robustness to arbitrary exposure counts or corruption patterns. It evaluates whether exposure-level rejection can reduce \ac{FLASH}-induced reconstruction deviation in the tested three-exposure, single-corruption setting.

\noindent \textbf{Additional mitigation directions.}
Two additional system-level approaches may reduce the reliability of \ac{FLASH}, although we do not evaluate them experimentally. First, capture-timing dither could introduce small randomized shifts in frame or bracket timing, making it harder for an open-loop pulsed source to repeatedly produce the same overlap pattern across exposure intervals. Such a method would require camera-driver or firmware support and may affect systems that expect stable capture timing. Second, cameras equipped with infrared illumination or automatic spotlight modes may reduce the relative effect of injected visible light by increasing scene illumination. This approach is hardware- and scene-dependent and should be treated as deployment-specific hardening rather than a general defense.

\noindent \textbf{Scope of the mitigation evaluation.}
Our defense evaluation considers three-exposure brackets in which corruption is injected only into the long exposure and at most one exposure is rejected. Although the rejection principle is intended to detect inconsistent exposure evidence more generally, we do not evaluate arbitrary numbers of exposures, corruption of other exposure positions, or simultaneous corruption of multiple exposures. The method also requires the corrupted exposure to exceed the anomaly threshold and at least one remaining exposure to preserve scene information. Abrupt benign lighting changes may also produce cross-bracket inconsistencies, and rejecting an exposure can reduce \ac{HDR} reconstruction quality because fewer measurements remain for fusion.

\begin{tcolorbox}[enhanced, colback=Cerulean!8, colframe=gray!70!black, toprule=0pt, bottomrule=0pt, rightrule=0pt, leftrule=2pt, boxsep=0pt, left=4pt, right=2pt, top=2pt, bottom=2pt]
\textbf{Defense Finding.}
In the controlled three-exposure night-only \ac{HDR} stress test, exposure-level rejection detects the corrupted long exposure in 756/756 cases with a 0.0\% clean false rejection rate and reduces median absolute luma deviation from 60.46\% to 12.60\%, a 79.16\% relative reduction. The result applies to the evaluated single-corruption setting; broader exposure configurations and corruption patterns remain unevaluated.
\end{tcolorbox}

%% file: 8-Discussion_and_Limitations.tex
\textbf{Sensor Fusion and Vehicle-Level Safety.}
\ac{FLASH} directly corrupts the camera stream without affecting independent radar or LiDAR sensing. In multi-sensor systems, these modalities may preserve evidence of pedestrians or obstacles and reduce risk, especially with redundant coverage, uncertainty estimation, sensor-health monitoring, and outlier rejection. However, fusion does not guarantee immunity: cameras remain important for lane markings, sign content, traffic-light state, object appearance, and other semantic information that radar or LiDAR may not recover. A system may also continue weighting a corrupted camera stream if it remains syntactically valid and triggers no failure indicator. Vehicle-level risk therefore depends on sensor coverage, fusion architecture, modality weighting, and failure handling, with camera-dominant systems and cases where secondary sensors provide incomplete evidence remaining more exposed.
Our system-level evaluation demonstrates degradation in the exported comma 3X camera stream and does not establish failure of object detection, sensor fusion, planning, or vehicle control. Evaluating whether radar or LiDAR fusion suppresses or propagates the effects of \ac{FLASH} requires an end-to-end multimodal vehicle platform and is left for future work.

\noindent\textbf{Attack Scope and Evaluation Limitations.}
\ac{FLASH} targets \emph{temporal} \ac{HDR} fusion and is not expected to transfer directly to single-exposure spatial \ac{HDR}. Architectures such as split-diode pixels, sub-pixel sensors, and programmable multi-tap sensors reduce the timing gap exploited by the attack~\cite{willassen20151280,SubPixel2018,MultiTap2025}. Our evaluation assumes non-contact line-of-sight deployment, low-to-moderate ambient lighting, and feasible emitter placement within safety and regulatory constraints.
The tested devices and environments are limited, and commercial cameras are evaluated as black-box pipelines because RAW brackets, fusion weights, and proprietary \ac{ISP} decisions are not exposed. We therefore rely on matched controls and a non-\ac{HDR} baseline to support the timing-dependent interpretation. The defense is evaluated in a controllable \ac{HDR} pipeline rather than inside proprietary camera firmware.

%% file: 9-Conclusion.tex
We introduced \ac{FLASH}, a physical attack that exploits the stable-illumination assumption of temporal \ac{HDR} fusion by creating inconsistent illumination across sequential exposures before downstream perception. Through simulation, matched optical controls, and physical evaluation across camera platforms, we show that this cross-bracket inconsistency produces pipeline-dependent failure modes, including darkening, overexposure, and local visibility degradation. The strongest physical results include extreme-darkening output rates of 50.0\% on the iPhone 16 Pro and 33.7\% on the Wyze Battery Cam Pro, while the controlled comma 3X case study shows up to a 90.8\% reduction in target-background CNR in the traffic-cone target region. In a controlled three-exposure night-only \ac{HDR} stress test, proof-of-concept exposure rejection reduces median absolute output-luma deviation by 79.16\%. These findings identify temporal \ac{HDR} fusion itself as a physical attack surface and show that camera pipelines should validate cross-exposure consistency before accepting exposure evidence for fusion.

%% file: 10-Ethical_appendix.tex
\noindent This research was conducted solely to assess theoretical vulnerabilities and improve camera imaging-pipeline safety; no actual interference in open traffic, private property, photography sessions, or public roadways was performed. Pedestrian examples were recorded with consent, and identifying facial regions were concealed and not published for anonymity.

%% file: 10-Open_science_Appendix.tex
\noindent To facilitate reproducibility and encourage further exploration of algorithmic vulnerabilities in \ac{ISP} pipelines, we provide the complete source code for our experimental framework. This release includes the full \ac{HDR}-Lab implementation and the Blender-based physical simulation parameter optimization environment presented in this work. \url{https://anonymous.4open.science/r/Lights-Camera-Attack-HDR-Manipulation-with-FLASH-Attacks-CB90/}.

%% file: 10-Appendix.tex
\section{Vulnerability Analysis of the \ac{HDR} Pipeline}
\label{Appendix:math_analysis}

\noindent \textbf{Global assumptions.}
We assume the camera operates in the linear response regime with fixed gain; RAW values are available before tone curves or clipping unless stated otherwise; the flash illuminates the same pixel set $R$ in every exposure; and the flash adds radiance $E_f>0$ without hard clipping except where noted. These assumptions isolate \ac{FLASH} from unrelated \ac{ISP} effects.

\noindent \textbf{Temporal HDR fusion model.}
Temporal \ac{HDR} captures a sequence of exposures with different integration times, typically short, medium, and long~\cite{debevec1997hdr,robertson1999high,mertens2007exposure}. For exposure $j$, the observed pixel value can be written as $I_j(x)=\Delta t_jE(x)$, where $E(x)$ is scene radiance and $\Delta t_j$ is exposure time. Under \ac{FLASH}, one exposure receives an additional radiance term, $I'_k(x)=\Delta t_k(E(x)+E_f(x))$. Because longer exposures have larger integration windows, they generally have a greater probability of overlapping a flash pulse. This can create cross-bracket inconsistency when one or more exposures receive substantially more injected energy than others, causing the exposure sequence to represent inconsistent scene illumination.

\noindent \textbf{Radiance bias.}
In weighted radiance reconstruction, per-pixel radiance is estimated from multiple exposures as
\[
\hat{E}(x)=
\frac{\sum_j w_j(x) I_j(x) / \Delta t_j}
     {\sum_j w_j(x)}.
\]
If exposure $k$ is flashed and not rejected, the estimate becomes
\[
\hat{E}'(x)
=
\hat{E}(x)
+
\frac{w_k(x)}{\sum_j w_j(x)}E_f(x),
\]
which biases the reconstructed \ac{HDR} map upward. The later tone-mapping stage may then compress the output as if the whole scene were brighter, producing darkened or extreme-darkening frames.

\noindent \textbf{Saturation and missing-data effect.}
If the flashed exposure saturates, its weight may be reduced or set to zero by saturation-aware fusion. In that case, the corrupted bracket becomes missing data, with $w_k(x)\approx0$. If the remaining exposures are underexposed, the fusion step has insufficient reliable information for the affected pixel region. This can reduce local detail, suppress target contrast, or force a dark fused value.

\noindent \textbf{Exposure-fusion failure.}
For exposure-fusion methods such as Mertens fusion, the per-pixel weight can be written as $W_j(x)=C_j(x)^{w_c}S_j(x)^{w_s}E_j(x)^{w_e}$, where $C_j$, $S_j$, and $E_j$ measure contrast, saturation, and well-exposedness~\cite{mertens2007exposure}. \ac{FLASH} can either overexpose the flashed bracket or make the non-flashed brackets appear inconsistent with it. In both cases, the normalized weights become unstable, and the fused image can lose target-background contrast.

\section{HDR-Lab Implementation Details}
    \label{app:hdrlab}

    \noindent The \ac{HDR}-Lab framework allows for a combinatorial evaluation of pipeline configurations under identical illumination conditions. It implements multiple canonical algorithms at each stage of the pipeline to enable controlled experimentation across alignment, merging, and tone mapping strategies.

    \noindent Table~\ref{tab:hdrlab_matrix} details the specific algorithms implemented within the framework. This modular design facilitates the evaluation of how specific standard techniques (such as shift-based alignment or global versus local tone mapping) react to adversarial illumination.

    \begin{table}[t]
    \centering
    \caption{\ac{HDR}-Lab Algorithm Comparison Matrix. The framework implements canonical algorithms at each stage of the pipeline to enable combinatorial evaluation of vulnerability to adversarial illumination.}
    \label{tab:hdrlab_matrix}
    \scriptsize
    \setlength{\tabcolsep}{3pt}
    \renewcommand{\arraystretch}{1.18}
    \begin{tabularx}{\columnwidth}{@{}p{0.20\columnwidth} p{0.27\columnwidth} X@{}}
    \toprule
    \rowcolor{white}
    \textbf{Pipeline Stage} & \textbf{Implemented Algorithms} & \textbf{Function / Characteristics} \\
    \midrule
    
    \rowcolor{gray!8}
    \textbf{Alignment}
    & MTB, Homography
    & Corrects for camera motion between exposure brackets. MTB is sensitive to global binary threshold shifts. \\
    
    \rowcolor{white}
    \textbf{Merging}
    & Robertson, Debevec, Mertens
    & Recovers response curves and fuses exposures. Mertens performs exposure fusion directly, bypassing radiance-map construction used by Robertson/Debevec. \\
    
    \rowcolor{gray!8}
    \textbf{Tone Mapping}
    & Mantiuk, Drago, Reinhard, Local
    & Compresses dynamic range for display. Global operators preserve saturation; local operators may introduce artifacts. \\
    
    \bottomrule
    \end{tabularx}
    \end{table}
    
\begin{figure}[h]
        \centering
        \includegraphics[width=\linewidth]{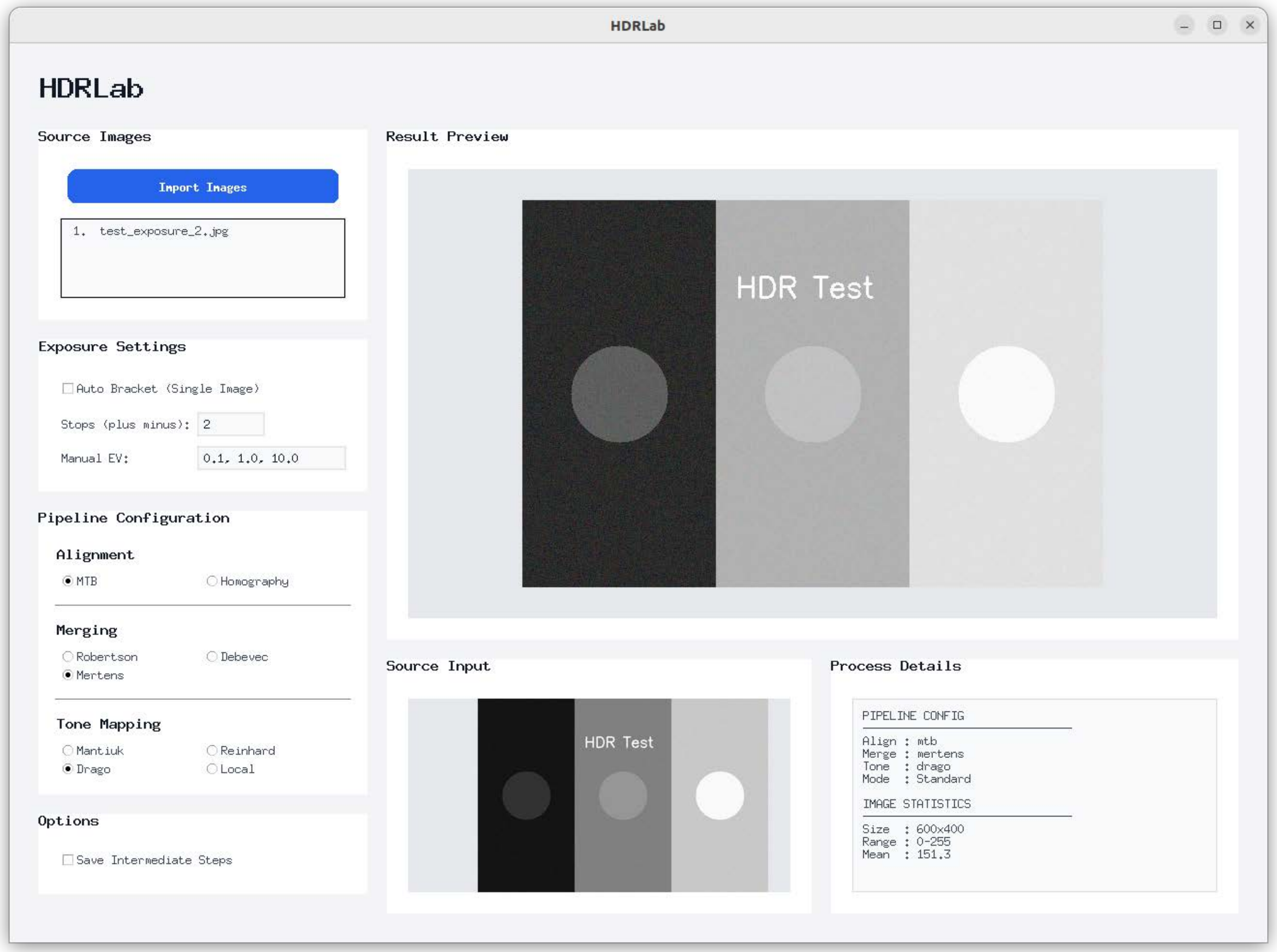}
        \caption{\ac{HDR}-Lab graphical interface used to configure and execute modular \ac{HDR} processing pipelines. The tool exposes swappable alignment, merging, and tone mapping stages and provides side-by-side visualization of source inputs, processed output, and per-run configuration details.}
        \label{fig:hdrlab_appendix}
    \end{figure}

    \noindent Figure~\ref{fig:hdrlab_appendix} illustrates the graphical interface used to configure these modular pipelines and visualize the impact of adversarial illumination on intermediate processing steps.
    
\section{Parameter Constraints and Interactions}
\label{app:flash_constraints_interactions}

The main methodology defines the attacker-selected configuration
\(\boldsymbol{a}\), operating condition \(\boldsymbol{s}\), component-wise
parameter bounds, received flash irradiance, and bracket-level injected
energy. This appendix specifies only the additional deployment constraints
used to construct the feasible set.

\noindent\textbf{Feasible set.}
Let \(\mathcal{B}\subset\mathbb{R}^{5}\) denote the Cartesian product of the
admissible intervals for the attacker-selected parameters defined in the
main methodology. For an operating condition \(\boldsymbol{s}\), we define
\[
\mathcal{C}(\boldsymbol{s})
=
\left\{
\boldsymbol{a}\in\mathcal{B}
\;\middle|\;
\begin{aligned}
&\ell_{\mathrm{LoS}}(\boldsymbol{a},\boldsymbol{s})=1,\\
&\delta_{\mathrm{clear}}(\boldsymbol{a},\boldsymbol{s})
    \geq \Delta_{\mathrm{safe}},\\
&c_{\mathrm{src}}(\boldsymbol{a})=1,\\
&c_{\mathrm{safe}}(\boldsymbol{a},\boldsymbol{s})=1
\end{aligned}
\right\}.
\]
Here, \(\ell_{\mathrm{LoS}}\) indicates whether the flashlight has an
unobstructed line of sight to the camera, and
\(\delta_{\mathrm{clear}}\) denotes the minimum clearance from
deployment-specific obstacles or protected regions. The condition
\(c_{\mathrm{src}}=1\) requires the source to support the selected power
and cadence under the fixed pulse-width configuration. The condition
\(c_{\mathrm{safe}}=1\) represents applicable deployment-specific safety
restrictions.

\noindent\textbf{Physical feasibility and attack effectiveness.}
Membership in \(\mathcal{C}(\boldsymbol{s})\) establishes only that a
configuration satisfies the placement, hardware, and safety constraints.
It does not require selective bracket contamination or a particular amount
of output degradation. These properties characterize attack effectiveness
rather than physical feasibility. Because the exact exposure windows and
relative phase are unavailable to the attacker, effectiveness is measured
from the resulting camera output in Section~\ref{sec:6-Experiments}.

\noindent\textbf{Motion-dependent feasibility.}
When the camera or source moves, let \(\boldsymbol{a}_k\) and
\(\boldsymbol{s}_k\) denote their values at frame \(k\). A configuration
remains feasible over an attack interval \(\mathcal{K}\) only if
\[
\boldsymbol{a}_k\in\mathcal{C}(\boldsymbol{s}_k),
\qquad
\forall k\in\mathcal{K}.
\]
Motion can change the source-to-camera geometry, received irradiance,
line-of-sight availability, and clearance. Motion itself does not change
the nominal cadence relationship unless the camera frame rate or the
source-camera clock relationship also changes.

\section{Metric Definitions} \label{app:metrics}
For each frame, we compute a Rec.709 luma proxy from decoded RGB values:
\(
Y' = 0.2126R' + 0.7152G' + 0.0722B',
\)
where \(R'\), \(G'\), and \(B'\) denote gamma-encoded color channels following the Rec.709 luma coefficients. We use \(Y'\) as an output-level brightness metric rather than as a calibrated photometric luminance measurement. Let \(\Omega\) denote an image region, such as the full frame, gray-card region, target region, or background region. For frame \(i\), the mean luma in region \(\Omega\) under condition \(c\) is
\(
\bar{Y}'_{i,\Omega,c}
=
\mathrm{mean}_{x\in\Omega}
\left(Y'_{i,c}(x)\right).
\)

For a test condition and its matched reference condition, the per-frame signed luma change is:
\[
\delta Y'_{i,\Omega}(\%)
=
100 \times
\frac{
\bar{Y}'_{i,\Omega,\mathrm{test}}
-
\bar{Y}'_{i,\Omega,\mathrm{ref}}
}
{
\bar{Y}'_{i,\Omega,\mathrm{ref}}
}.
\]

For parameter-sweep trend analysis, we also use the decoded luma-unit change
\(
\Delta Y'_{i,\Omega}
=
\bar{Y}'_{i,\Omega,\mathrm{test}}
-
\bar{Y}'_{i,\Omega,\mathrm{ref}}.
\)
We report the median signed luma-unit change \(\widetilde{\Delta Y'}_{\Omega}\) over matched frames for the physical parameter sweeps. This luma-unit metric is used only to summarize sweep trends; the default cross-device and system-level failure metrics remain percent-normalized.

For the default cross-device and system-level comparisons, we report the mean, median, and minimum signed luma change over frames
\(\bigl(\overline{\delta Y'}_{\Omega},
\widetilde{\delta Y'}_{\Omega},
\delta Y'_{\min,\Omega}\bigr)\),
along with the severe-darkening and extreme-darkening rates:
\[
\begin{aligned}
r_{\mathrm{sev},\Omega}
&=
\frac{1}{N}
\sum_{i=1}^{N}
\mathbb{1}\!\left[\delta Y'_{i,\Omega}\le -50\%\right],\\
r_{\mathrm{ext},\Omega}
&=
\frac{1}{N}
\sum_{i=1}^{N}
\mathbb{1}\!\left[\delta Y'_{i,\Omega}\le -80\%\right].
\end{aligned}
\]
Here, \(N\) is the number of matched frames used in the comparison. Thus, severe darkening is the percentage of frames with at least a 50\% luma reduction relative to the matched reference, and extreme-darkening rate is the percentage of frames with at least an 80\% luma reduction.

We also compute target-background contrast-to-noise ratio using the same luma proxy. Let \(\Omega_T\) and \(\Omega_B\) denote the target and background regions. For frame \(i\) under condition \(c\), the per-frame CNR is:
\[
\mathrm{CNR}_{i,c}
=
\frac{
\left|
\bar{Y}'_{i,\Omega_T,c}
-
\bar{Y}'_{i,\Omega_B,c}
\right|
}
{
\sqrt{
\sigma^2_{i,\Omega_T,c}
+
\sigma^2_{i,\Omega_B,c}
}
},
\]
where \(\bar{Y}'_{i,\Omega,c}\) is the mean Rec.709 luma proxy in region \(\Omega\), and \(\sigma^2_{i,\Omega,c}\) is the corresponding regional luma variance. For a test condition and its matched reference condition, the per-frame signed CNR change is
\[
\delta\mathrm{CNR}_{i}(\%)
=
100 \times
\frac{
\mathrm{CNR}_{i,\mathrm{test}}
-
\mathrm{CNR}_{i,\mathrm{ref}}
}
{
\mathrm{CNR}_{i,\mathrm{ref}}
}.
\]
We report summary statistics of \(\delta\mathrm{CNR}_{i}\) over frames for comparisons where target and background regions are defined. This metric captures changes in target-background separability under the same decoded-output measurement model used for the luma-change metrics.
\section{Simulation Environment Details}
\label{app:sim_imp_details}

\noindent The simulation is conducted in Blender within a closed indoor cube of size \SI{50}{m} $\times$ \SI{50}{m} $\times$ \SI{50}{m}. The cube origin is placed at $z=\SI{25}{m}$ so that the floor lies on the world $z=0$ plane. The interior face normals are flipped to support internal ray tracing, and the ceiling face is removed to admit overhead ambient illumination. The interior surfaces use a checker texture with scale 10.0 and a noise texture with scale 60.0 and detail 15.0, multiplied into the base color to introduce fine spatial variation. All surfaces use a Principled BSDF with specular IOR level 0.0 and roughness 1.0, which suppresses mirror-like reflections and keeps the measurements dominated by diffuse irradiance. Atmospheric effects are modeled with a world volume that applies volume scatter with density 0.01 and anisotropy $-1.0$. Ambient illumination is provided by a \SI{50}{m} $\times$ \SI{50}{m} area light positioned at a height of \SI{50}{m}. We use two ambient source-power settings, $P_a^{\mathrm{sim}}=\SI{500}{kW}$ for daytime and $P_a^{\mathrm{sim}}=\SI{20}{kW}$ for nighttime. These are simulation source-power settings and are not treated as equivalent to the lux measurements used in the physical evaluation. The attack source is modeled as a circular emitter with diameter \SI{0.1}{m}, and shadow visibility is disabled to avoid self-occlusion artifacts from the source geometry. The evaluated optical power levels are \SIlist{2;4;8;12;20}{W}, corresponding to the representative commercial source classes summarized in Table~\ref{tab:flashlight_tiers}.

\begin{table}[t]
    \centering
    \caption{Simulated optical power levels and representative commercial source classes.}
    \label{tab:flashlight_tiers}
    \small
    \begin{tabular}{@{}lc@{}}
        \toprule
        \textbf{Representative Source Class} &
        \textbf{Optical Power} \\
        \midrule
        \rowcolor{gray!8}
        Low-power strobe & \SI{2}{W} \\
        \rowcolor{white}
        Everyday-carry flashlight & \SI{4}{W} \\
        \rowcolor{gray!8}
        Security flashlight & \SI{8}{W} \\
        \rowcolor{white}
        High-output flashlight & \SI{12}{W} \\
        \rowcolor{gray!8}
        Searchlight-class source & \SI{20}{W} \\
        \bottomrule
    \end{tabular}
\end{table}

The spatial sweep evaluates distances from \SI{2}{m} to \SI{20}{m} in \SI{2}{m} increments and horizontal angles from $-45^{\circ}$ to $45^{\circ}$. The height sweep fixes the distance at \SI{4.0}{m} and the angle at $0^{\circ}$, then evaluates source heights of \SIlist{0.5;1.5;3.0}{m}. The power sweep uses the same \SI{4.0}{m} distance, $0^{\circ}$ angle, and \SI{1.5}{m} source height while varying source power across the five evaluated levels.

\section{Component-Level Physical Setup}
\label{app:physical_setup}

\begin{figure}[H]
    \centering
    \includegraphics[width=\columnwidth]{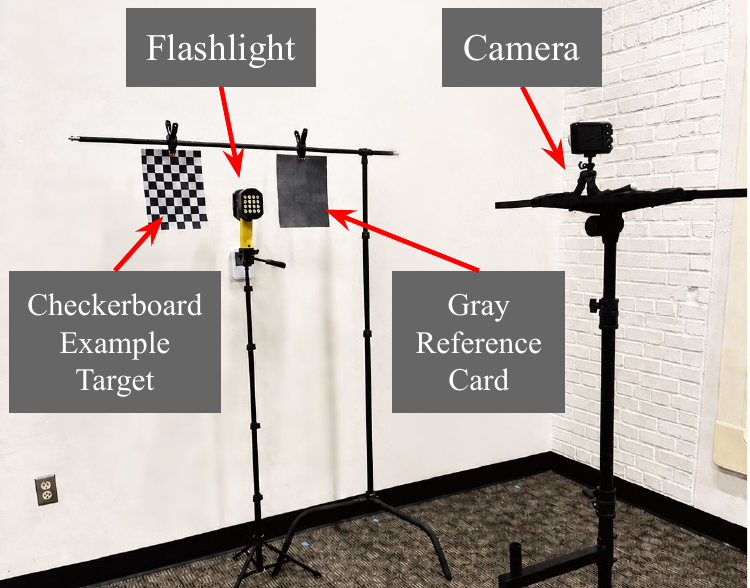}
    \caption{Component-level physical setup used for the \ac{FLASH} evaluation. The setup includes the target board, checkerboard target, gray reference card, flash source, and camera under test.}
    \label{fig:physical_setup}
\end{figure}

Figure~\ref{fig:physical_setup} shows the fixed component-level setup used across the physical camera experiments.

\begin{figure*}[t]
    \centering
    \includegraphics[width=0.875\textwidth]{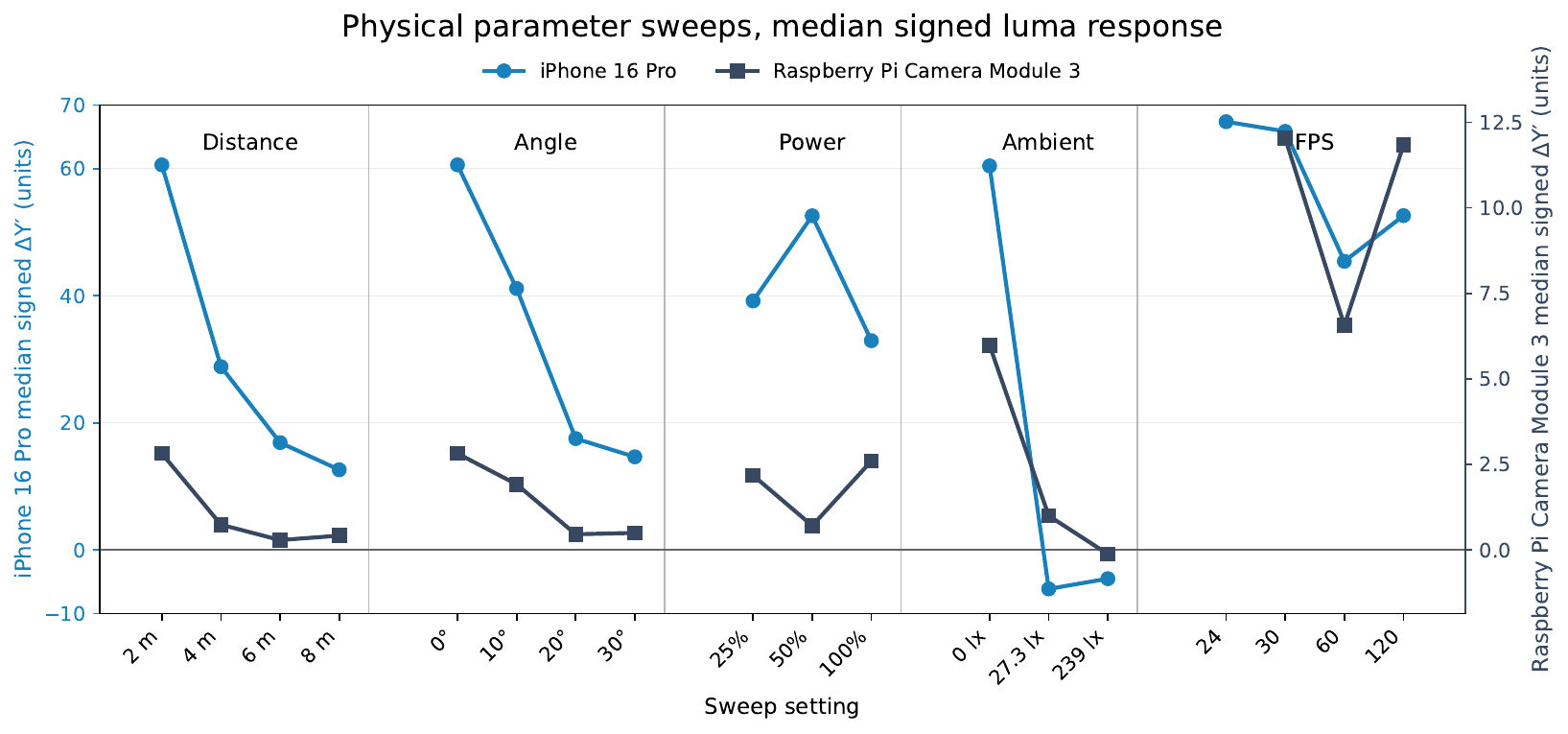}
    \caption{Complete physical parameter sweep results for the iPhone 16 Pro and Raspberry Pi Camera Module 3. Values report median signed decoded luma-unit change for matched clean-versus-FLASH trials. Separate y-axis scales preserve the within-device trends.}
    \label{fig:physical_sweeps}
\end{figure*}

We conduct the component-level physical evaluation in a fixed indoor setup. The camera under test is mounted on a tripod at \(z_c=\SI{1.5}{m}\), and the target board is placed \SI{2}{m} from the camera. The flash source is mounted on a separate tripod at \(z_f=\SI{1.5}{m}\) and aimed toward the camera and target region. The scene includes a checkerboard target and a gray reference card. Ambient condition \(A\) and control-light illuminance are measured at the camera plane in lux, and floor markings maintain consistent camera, target, distance, and angle positions across devices.
All physical cameras are treated as black-box imaging pipelines, and we record the final image or video output using the available or default capture mode. The default physical setup uses a manufacturer-rated \(P_f=\SI{2}{W}\) commodity flashlight at \(d_{fc}=\SI{2}{m}\) to provide sufficient injected illumination. The simulation default instead uses \(P_f=\SI{12}{W}\) at \(d_{fc}=\SI{4}{m}\). This provides a controlled mid-strength simulation condition without trivially saturating the closest setting with the strongest \SI{20}{W} source.
\section{Physical Operational Robustness Details}
\label{app:physical_robustness}

The saved physical-camera dataset contains 94 videos: 28 default videos, 28 distance and angle sweep videos, 12 source-setting sweep videos, 12 ambient-illuminance sweep videos, and 14 frame-rate sensitivity videos. Figure~\ref{fig:physical_sweeps} reports the median signed decoded luma-unit response across the physical sweep settings.

\noindent \textit{Distance sweep (\(d_{fc}\)).}
The distance sweep shows the strongest response at the closest tested range. On the iPhone 16 Pro, the median \(\Delta Y'\) decreases from \(+60.60\) units at \SI{2}{m} to \(+28.85\), \(+16.90\), and \(+12.62\) units at \SIlist{4;6;8}{m}. The Raspberry Pi Camera Module 3 shows the same attenuation with smaller magnitude, decreasing from \(+2.82\) units at \SI{2}{m} to \(+0.74\), \(+0.29\), and \(+0.42\) units at \SIlist{4;6;8}{m}. This confirms that the received flash contribution weakens as \(d_{fc}\) increases.

\noindent \textit{Angle sweep (\(\theta\)).}
The angle sweep peaks near the optical axis. On the iPhone 16 Pro, the median \(\Delta Y'\) decreases from \(+60.60\) units at \(0^{\circ}\) to \(+41.15\), \(+17.53\), and \(+14.66\) units at \(10^{\circ}\), \(20^{\circ}\), and \(30^{\circ}\). The Raspberry Pi Camera Module 3 decreases from \(+2.82\) units at \(0^{\circ}\) to \(+1.92\), \(+0.46\), and \(+0.50\) units. This matches the simulation results showing that off-axis placement reduces the injected light reaching the camera.

\noindent \textit{Source-setting sweep (\(P_f\)).}
The source-setting sweep is non-monotonic, which is expected for black-box camera pipelines with automatic exposure and tone mapping. On the iPhone 16 Pro, the median \(\Delta Y'\) is \(+39.19\), \(+52.59\), and \(+32.94\) units at 25\%, 50\%, and 100\% settings. On the Raspberry Pi Camera Module 3, the corresponding values are \(+2.18\), \(+0.71\), and \(+2.60\) units. Thus, a higher \(P_f\) setting does not always produce a larger final luma shift because the camera pipeline can compensate differently across settings.

\noindent \textit{Ambient-illuminance sweep (\(A\)).}
Increasing measured ambient illuminance suppresses the low-light response. On the iPhone 16 Pro, the median \(\Delta Y'\) changes from \(+60.44\) units at \SI{0}{lux} to \(-6.12\) units at \SI{27.3}{lux} and \(-4.51\) units at \SI{239}{lux}. On the Raspberry Pi Camera Module 3, the response decreases from \(+5.98\) units at \SI{0}{lux} to \(+1.00\) and \(-0.12\) units at \SI{27.3}{lux} and \SI{239}{lux}. These results show that increasing \(A\) reduces the relative contribution of the injected flash.

\noindent \textit{Frame-rate sensitivity (\(f_{\mathrm{fps}}\)).}
The frame-rate sweep shows that capture mode affects the response, but the effect is not confined to one frame rate. On the iPhone 16 Pro, the median \(\Delta Y'\) is \(+67.39\), \(+65.87\), \(+45.43\), and \(+52.61\) units at \SIlist{24;30;60;120}{fps}. On the Raspberry Pi Camera Module 3, the response is \(+12.03\), \(+6.56\), and \(+11.83\) units at \SIlist{30;60;120}{fps}. These results show that timing behavior and capture mode shape the \ac{FLASH} response, while the attack effect remains measurable across the tested \(f_{\mathrm{fps}}\) settings.
\section{Legal Implications of Blinding \ac{AV} and Surveillance Cameras} \label{Appendix:legal_analysis}
\noindent Deliberate optical interference with \ac{AV} sensors, including high-intensity flashlights or lasers aimed at cameras or LiDAR, is likely covered by existing criminal statutes. Federal law treats willful acts that “damage, disable, or tamper with” a motor vehicle in a way that endangers human life as a felony punishable by up to twenty years’ imprisonment (18 U.S.C. § 33)\cite{uscode33,doj1426}. State laws add related prohibitions: Texas may classify intentional sensor blinding as criminal mischief\cite{texas28,findlaw_tx28}, while California forbids directing a laser beam into a moving vehicle with intent to harass or annoy\cite{ca41727,ca41725}. Related civil claims may include negligence, trespass to chattels, vandalism, repair costs, medical costs, and punitive damages. Similar rules apply to interference with security or traffic cameras, including vandalism, criminal trespass, and public nuisance theories\cite{ny14514}.

\section{Publication Disclaimer}
Reference herein to any specific commercial company, product, process, or service by trade name, trademark, manufacturer, or otherwise, does not necessarily constitute or imply its endorsement, recommendation, or favoring by the United States Government or the Department of the Army (DoA). The opinions of the authors expressed herein do not necessarily state or reflect those of the United States Government or the DoA and shall not be used for advertising or product endorsement purposes.